\documentclass[prl,twocolumn,amssymb,superscriptaddress]{revtex4-2}
\usepackage{amsmath}
\usepackage{amssymb}
\usepackage{amsthm}
\usepackage{amsfonts}
\usepackage[version=4]{mhchem}
\usepackage{listings}
\usepackage{enumerate}
\usepackage{latexsym}
\usepackage{psfrag}
\usepackage{bm}
\usepackage[all]{xy}
\usepackage{graphicx}
\usepackage[pdftex,colorlinks]{hyperref}
\usepackage{color}
\usepackage{times}
\usepackage{array}
\usepackage{placeins}

\begin{document}

\title{Doping evolution of superconductivity, charge order and band topology in hole-doped topological kagome superconductors Cs(V$_{1-x}$Ti$_x$)$_3$Sb$_5$}
\author{Yixuan Liu}
\thanks{These authors contributed equally.}
\author{Yuan Wang}
\thanks{These authors contributed equally.}
\author{Yongqing Cai}
\thanks{These authors contributed equally.}
\author{Zhanyang Hao}
\thanks{These authors contributed equally.}

\author{Xiao-Ming Ma}
\author{Le Wang}
\author{Cai Liu}
\affiliation{Shenzhen Institute for Quantum Science and Engineering, and Department of Physics, Southern University of Science and Technology, Shenzhen
  518055, China}
\affiliation{International Quantum Academy, and Shenzhen Branch, Hefei National
  Laboratory, Futian District, Shenzhen 518048, China}
\author{Jian Chen}
\author{Liang Zhou}
\author{Jinhua Wang}
\author{Shanming Wang}
\author{Hongtao He}
\affiliation{Shenzhen Institute for Quantum Science and Engineering, and Department of Physics, Southern University of Science and Technology, Shenzhen
  518055, China}
\author{Yi Liu}
\author{Shengtao Cui}
\affiliation{National Synchrotron Radiation Laboratory, University of Science
  and Technology of China, Hefei 230029, China}
\author{Jianfeng Wang}
\email{wangjf06@buaa.edu.cn}
\affiliation{School of Physics, Beihang University, Beijing 100191, China}
\affiliation{Beijing Computational Science Research Center, Beijing 100193,
  China}
\author{Bing Huang}
\affiliation{Beijing Computational Science Research Center, Beijing 100193,
  China}
\author{Chaoyu Chen}
\email{chency@sustech.edu.cn}
\affiliation{Shenzhen Institute for Quantum Science and Engineering, and Department of Physics, Southern University of Science and Technology, Shenzhen
  518055, China}
\affiliation{International Quantum Academy, and Shenzhen Branch, Hefei National
  Laboratory, Futian District, Shenzhen 518048, China}
\author{Jia-Wei Mei}
\email{meijw@sustech.edu.cn}
\affiliation{Shenzhen Institute for Quantum Science and Engineering, and Department of Physics, Southern University of Science and Technology, Shenzhen
  518055, China}
\affiliation{International Quantum Academy, and Shenzhen Branch, Hefei National
  Laboratory, Futian District, Shenzhen 518048, China}
\affiliation{Shenzhen Key Laboratory of Advanced Quantum Functional Materials
  and Devices, Southern University of Science and Technology, Shenzhen 518055, China}

 \date{\today}

 \begin{abstract}
The newly discovered Kagome superconductors $A$V$_3$Sb$_5$ ($A$ = K, Rb, Cs) exhibit superconductivity, charge order, and band topology simultaneously. To explore the intricate interplay between the superconducting and charge orders, we investigate the doping evolution of superconductivity, charge-density-wave (CDW) order and band topology in doped topological kagome superconductors Cs(V$_{1-x}$Ti$_x$)$_3$Sb$_5$ where the Ti-dopant introduces hole-like charge carriers.  Despite the absence of the CDW phase transition in doped compounds even for the lowest doping level of $x=0.047$, the superconductivity survives in all doped samples with enhanced critical temperatures. The high-resolution angle-resolved photoemission spectroscopy (ARPES) measurements reveal that the Ti-dopant in the kagome plane lowers the chemical potential, pushing the van Hove singularity (VHS) at $M$ point above the Fermi level. First-principle simulations corroborate the doping evolution of the band structure observed in ARPES, and affirm that the CDW instability does not occur once the VHS is pushed above the Fermi level, explaining the absence of the CDW ordering in our doped samples Cs(V$_{1-x}$Ti$_x$)$_3$Sb$_5$. Our results demonstrate a competition between the CDW and superconducting orders in the kagome-metal superconductor CsV$_3$Sb$_5$, although the superconductivity is likely inconsequential of the CDW order. 
\end{abstract}
\maketitle
The transition-metal kagome compounds host novel correlated and topological electronic states and have been catching lots of attention in recent years~\cite{Isakov2006,Guo2009,Tang2011,Nakatsuji2015,Liu2018,Ye2018,LiuD.2019,Yin2020,Ortiz2019,Ortiz2020,Ortiz2021}. Due to the geometrical frustration for electron hoppings, the simple kagome-lattice band structure possesses Dirac cones, flat band, and saddle point with van Hove singularities (VHS) in the electronic density of states (DOS), facilitating the emergence of various electronic  orders~\cite{Isakov2006,Johannes2008,Guo2009,Tang2011,Wang2013,Kiesel2013}. Among the kagome metal compounds, the charge-density-wave (CDW) charge order and superconductivity arise successively at around 78-103~K and 0.92-2.5~K, respectively, and intertwine with each other at low temperatures in the $A$V$_3$Sb$_5$ ($A$=K, Rb, Cs) family~\cite{Ortiz2019,Ortiz2020,Ortiz2021,Chen2021b,Zhao2021}, provoking hot topics of recent interests. The angle-resolved-photoemission spectroscopy (ARPES) also reveals a band topology with multiple Dirac lines and loops and a rich nature of VHSs close to the Fermi level in $A$V$_3$Sb$_5$~\cite{Ortiz2020,Luo2021,Li2021,Kang2021,Lou2021,Cho2021,Hu2021,Hu2021a,Cai2021}.

\begin{figure*}[t]
  \centering
  \includegraphics[width=2\columnwidth]{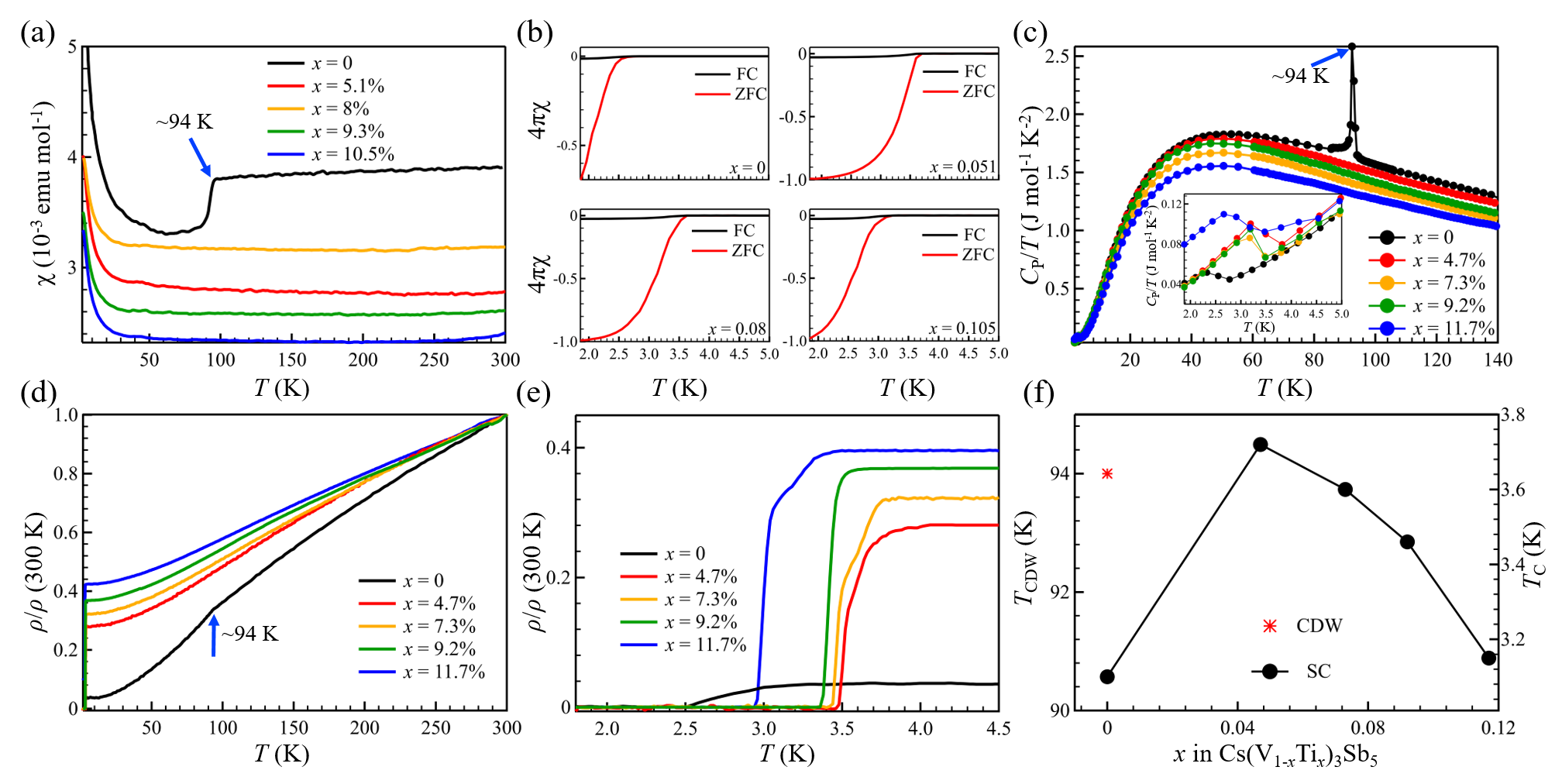}
  \caption{Thermodynamic and transport properties of
    Cs(V$_{1-x}$Ti$_x$)$_3$Sb$_5$. (a) Temperature-dependent uniform magnetic
    susceptibilities $\chi=M/H$ with the $c$-axis directional field $H=10^4$~Oe.
    (b) The zero-field-cool (ZFC) and field-cool (FC) magnetic susceptibility
    data under $H = 20$~Oe.  (c) Temperature-dependent specific heat coefficient
    $C_p/T$ at zero field. The inset is the zoom-in low temperature data. (d)
    Temperature-dependent resistivity $\rho$ with respect to the 300~K result.
    (e) The zoom-in low temperature data of (d). (f) The doping dependent CDW and superconducting transition temperatures.}
  \label{fig:figure1}
\end{figure*}

The CDW order in $A$V$_3$Sb$_5$ ($A$=K, Rb, Cs) is a $2\times2\times2$
three-dimensional one  with unconventional
properties~\cite{Li2021,Luo2021,Luo2021a,Liang2021,Jiang2021,Feng2021,Denner2021}.
Although the muon spin spectroscopy measurement does not detect any local moment for the vanadium ions~\cite{Kenney2021}, a giant anomalous Hall conductivity was observed~\cite{Yang2020,Yu2021a} below the CDW transition temperature, implying a non-trivial chirality with the time-reversal symmetry breaking~\cite{Jiang2021,Feng2021,Denner2021}. The scanning tunneling microscope/spectroscopy reveals nematic/sematic order and paring-density-wave order in CsV$_3$Sb$_5$~\cite{Zhao2021,Chen2021b}, by interesting analogy with intertwined orders in the high-temperature cuprate superconductors~\cite{Fradkin2015,Dai2018}. The CDW order  can be completely suppressed under low pressures with emerging the superconducting domes~\cite{Chen2021,Du2021,Yu2021}, implying  a competition between the CDW order and the superconductivity~\cite{Wilson2021}.  The fragility of CDW against pressures suggests that the electronic structure determines its nature, however, no ARPES measurement of the band structure in response to external perturbations has been reported.

In this letter, we substitute vanadium with titanium in CsV$_3$Sb$_5$ to
modulate the chemical potential and study the doping evolution of
superconductivity, charge order and band topology in
Cs(V$_{1-x}$Ti$_x$)$_3$Sb$_5$. We do not observe the CDW phase transition in the
doped compounds even for the doping level of $x=0.047$ in the transport and
thermodynamic measurements. Meanwhile, the superconductivity remains with an
enhanced temperature in all doped samples. From the ARPES measurements, we
reveal that the overall pattern of band structures remains unchanged upon
doping, but the Fermi level lowers down and hence the VHS1 at the $M$ point
(Fig.~\ref{fig:figure3}) rises up in
Cs(V$_{1-x}$Ti$_x$)$_3$Sb$_5$. Our first-principles simulations confirm the band
evolution upon doping and affirm that the CDW instability in the parent
compound~\cite{Tan2021} does not occur once VHS1 is pushed above the Fermi
level, accounting for the absence of CDW  in the doped samples. Except for the
parent compound, the superconducting temperature $T_c$ decreases with increasing
the doping levels, coinciding with the DOS of the Sb-derived electron pocket at
$\Gamma$ point. The parent compound has a larger DOS of the Sb electron pocket, but a lower $T_c$ than the doped samples, indicating a competition between the CDW order and the superconductivity.


\begin{figure*}[t]
  \centering
  \includegraphics[width=2\columnwidth]{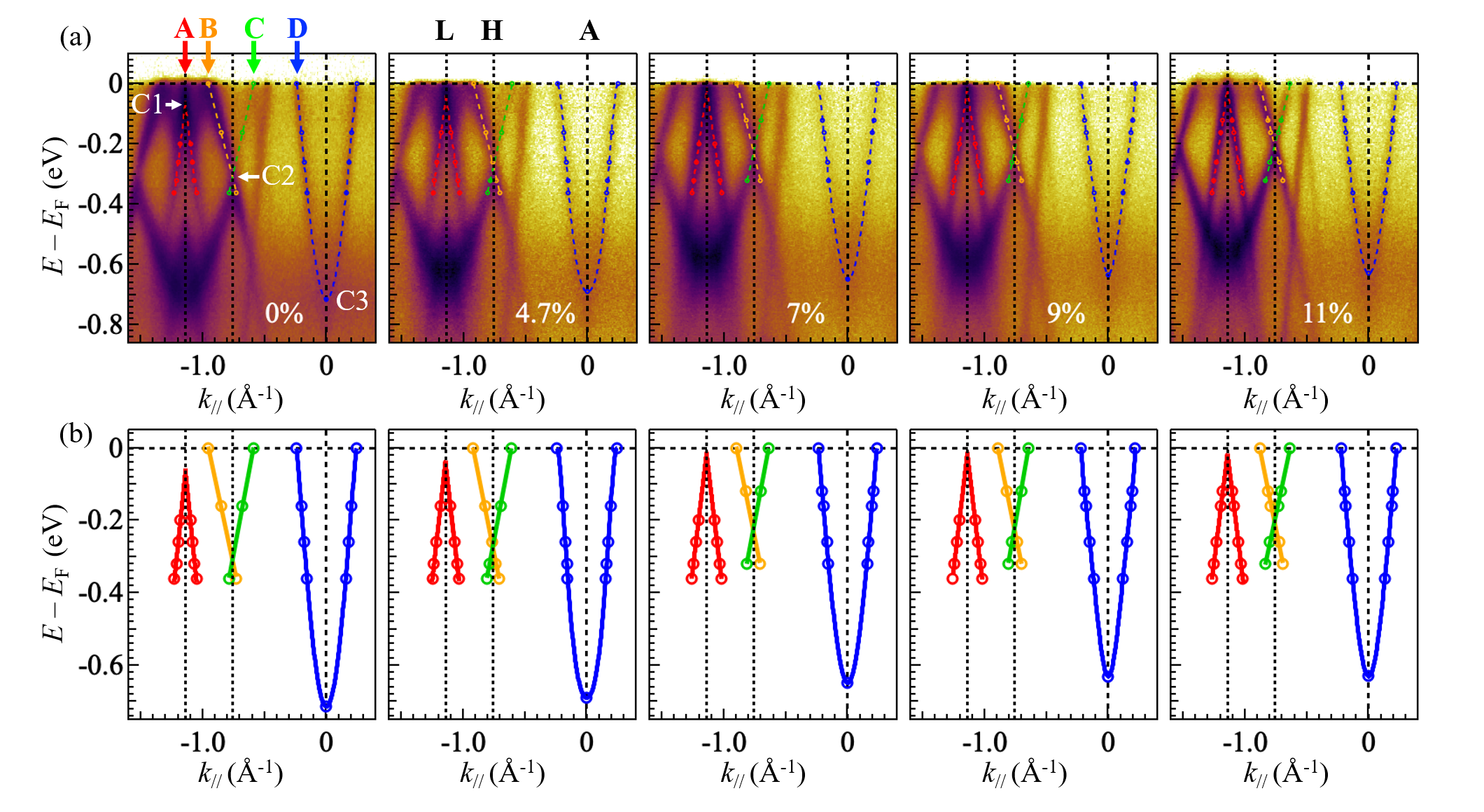}
  \caption{The doping evolution of the electronic structures of Cs(V$_{1-x}$Ti$_x$)$_3$Sb$_5$. (a) Band structures measured along the A-H-L directions with 31 eV photons at 17 K with different doping levels. (b) The corresponding schematic band dispersions extracted from (a).}
  \label{fig:figure2}
\end{figure*}

Cs(V$_{1-x}$Ti$_x$)$_3$Sb$_5$ single crystals with the titanium doping were synthesized by the
self-flux method with the starting ratio ${\rm Cs}: {\rm V}: {\rm Ti}: {\rm Sb} = 2: 1:
x_{\rm Ti}
: 6$ for $0\leq x_{\rm Ti}\leq 0.5$~\cite{SM}.  The chemical composition of
 CS(V$_{1-x}$Ti$_x$)$_3$Sb$_5$ was identified by Energy Dispersive X-Ray
 Spectroscopy (EDX, Phenom ProX).
Fig.~\ref{fig:figure1} summarizes the thermodynamic and transport properties
of Cs(V$_{1-x}$Ti$_x$)$_3$Sb$_5$ with different doping levels $x$. Titanium sits
on the left next to vanadium, and the substitution reduces the total number of
charge carriers. The lowest doping concentration with $x=0.047$ of our samples
already completely suppresses the CDW transition. However, the superconducting
state at low temperatures survives at all doping levels ($0.047\leq x\leq0.117$)
with enhanced critical temperatures. In Fig.~\ref{fig:figure1}(a), we compare
the temperature-dependent magnetic susceptibilities of the parent compound
CsV$_3$Sb$_5$ and the Ti-doped ones Cs(V$_{1-x}$Ti$_x$)$_3$Sb$_5$.  A sudden
drop of the magnetization in CsV$_3$Sb$_5$ appears at 94~K, indicating the CDW
transition~\cite{Ortiz2019}. For all Ti-doped samples, no CDW anomaly occurs,
implying the absence of the CDW transition. We observe increasingly
weakened magnetization signature at both high and low temperatures with the
increasing doping levels $x$, suggesting a reduction of DOS of itinerary
electrons or local correlations for vanadium 3$d$ electrons upon doping. The
complete suppression of CDW is also revealed in the heat capacity and the
resistivity measurements in Figs.~\ref{fig:figure1}(c) and (d)-(e). All the  samples of Cs(V$_{1-x}$Ti$_x$)$_3$Sb$_5$ display the superconducting transition at low temperatures in the magnetization (Fig.~\ref{fig:figure1}(b)), heat capacity (Fig.~\ref{fig:figure1}(c)) and resistivity (Fig.~\ref{fig:figure1}(d)-(e)) measurements. Fig.~\ref{fig:figure1}(f) is the doping-dependent transition temperatures of the CDW order and the superconductivity.

High-resolution ARPES measurements are performed to investigate the band structure evolution of the Cs(V$_{1-x}$Ti$_x$)$_3$Sb$_5$ with different doping level. As detailed photon energy dependent ARPES measurements have been performed~\citep{Cai2021}, the value of k$_z$ can be determined precisely. Fig.~\ref{fig:figure2}(a) shows band structures measured at 17 K of different doping level with 31 eV photons, which correspond to the A-H-L plane (k$_z$ = $\pi$). As we can see, an electron-like band around A, several Dirac cone-like dispersions along the A-H-L direction can be clearly resolved in this system. To focus on how these bands evolve with different doping, we extract the their band dispersions by fitting the momentum distribution curves (MDCs), marked by A-D in different colors, and then draw them in Fig.~\ref{fig:figure2}(b) separately. The overall band structures shift upwards with different doping level monotonously, in accordance with the fact that hole doping pushes the Fermi level ($E_{\rm F}$) downwards. By substituting 11$\%$ percent of V atoms with Ti atoms, the band structure can be shifted up about 100 meV. Our results prove that the substitution of Ti is an efficient way to tune the Fermi level of CsV$_3$Sb$_5$ system.

\begin{figure*}[t]
  \centering
  \includegraphics[width=2\columnwidth]{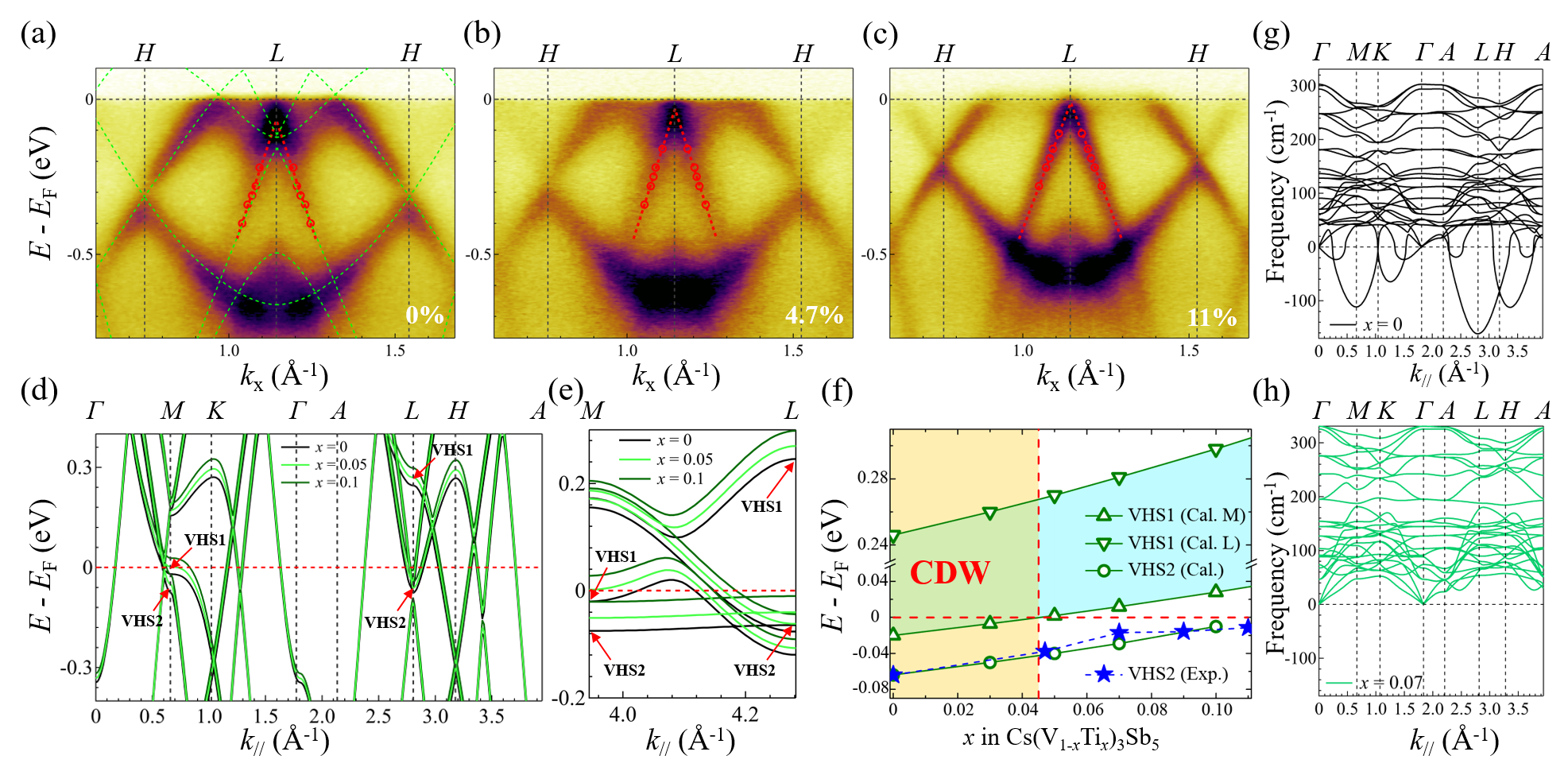}
  \caption{Doping evolution of VHSs at the $M/L$ points in
    Cs(V$_{1-x}$Ti$_x$)$_3$Sb$_5$. (a)-(c) ARPES measurements along the H-L-H
    direction with selected doping levels. In (a) the DFT calculated band
    structure (green dashed lines) is also over-plotted. The DFT calculated
    projected bulk band structure along $\Gamma$-M-K-$\Gamma$ and A-L-H-A
    directions in (d) and along the M-L direction in (e). (f) The comparison between the calculated and experimental results on the energy shift of VHSs with doping. (g-h) The calculated phonon dispersion relations with doping level $x = 0$ (g) and $x= 0.07$ (h).}
  \label{fig:figure3}
\end{figure*}

As shown in Fig.~\ref{fig:figure1}, there is no CDW anomaly for all the Ti-doped CsV$_3$Sb$_5$ compounds. Since the VHSs at the M point near the Fermi level play crucial roles in the CDW instability, it is of importance to figure out how the VHSs near the Fermi level evolve with different doping. We select three typical doping levels (x = 0, 0.047 and 0.11) and analyse their detailed band structures around the L point, as shown in Figs.~\ref{fig:figure3}(a-c). According to the previous ARPES study~\cite{Cai2021}, the ARPES spectral intensity is dominated by contribution from only the top two layers, resulting in the lack of clear k$_z$ dispersion along M-L and the missing information of the VHSs around the M point. To have a fully clear understanding on the evolutions of band structure and VHSs upon doping, first-principles simulations within density-functional theory (DFT) are performed here~\cite{SM}. As revealed by ARPES measurements in Fig.~\ref{fig:figure2}, except for the overall upward shift, all the features of band structure remains almost unchanged under doping, meaning that the Ti-dopant doesn't break any crystal symmetry of CsV$_3$Sb$_5$. Therefore, we employ a virtual crystal approximation in the simulations of Ti-substituted V in CsV$_3$Sb$_5$, which is qualitatively consistent with the results calculated using the supercell of Cs(V$_{1-x}$Ti$_x$)$_3$Sb$_5$~\cite{SM}. Fig.~\ref{fig:figure3}(a) demonstrates that the DFT calculated band structure can well fit the ARPES measurements of CsV$_3$Sb$_5$. As shown in Fig.~\ref{fig:figure3}(d), upon doping of Ti, the overall band structures are shifted upwards, consistent with the hole-doped effect revealed by ARPES. Now we focus on the evolutions of VHSs. Around the Fermi level, there are two VHSs (VHS1 and VHS2) at both the M and L points, as marked by red arrows in Fig.~\ref{fig:figure3}(d). It is noted that the VHS1 has a large $k_z$ dispersion along M-L, and it crosses the Fermi level from the M (below $E_{\rm F}$) to L (above $E_{\rm F}$) points, as shown in Fig.~\ref{fig:figure3}(e). While the VHS2 has almost no $k_z$ dispersion along M-L. In addition, the VHS1 has a higher density of states (DOS) near $E_{\rm F}$, which accounts for the CDW instability as will be demonstrated below.

Around the L point, the VHS1 cannot be measured by ARPES due to $E(\rm VHS1)>$$E_{\rm F}$, but the VHS2 is observed under all our doping levels, as shown in Figs.~\ref{fig:figure3}(a-c). We extract the band dispersions around VHS2 [red circles in Figs.~\ref{fig:figure3}(a-c)] and fit them with red dashed lines. Thus the exact positions of VHS2 can be obtained quantitatively. In Fig.~\ref{fig:figure3}(f), we summarize the calculated and experimental energy shifts of these two VHSs under different doping levels where both the energies of VHS1 at M and L are plotted here. In general, all the energies of VHSs are shifted upward with respect to $E_{\rm F}$ as increasing doping level. The calculated energy evolution of VHS2 [green circles in Fig.~\ref{fig:figure3}(f)] is in good agreement with the experimental results [blue stars in Fig.~\ref{fig:figure3}(f)]. Focusing on the VHS1, its $k_z$ dispersion along M-L allows the VHS1 to be located at any energy position between the energies of the M and L points, which has been plotted by the cyan region in Fig.~\ref{fig:figure3}(f). In other words, the VHS1 can be in principle located at any position of the cyan region except for some energy gaps caused by the coupling between VHS1 and other bands. It is noted that the CDW instability is mainly induced by the VHS at $E_{\rm F}$. Therefore, that the Fermi energy is included in the range of VHS1 energy between the M and L points is a necessary condition for the formation of CDW. As shown in Fig.~\ref{fig:figure3}(f), for the parent compound CsV$_3$Sb$_5$, the VHS1 at M (L) is below (above) $E_{\rm F}$. With hole doping, the VHS1 at M shifts upward and is pushed above the Fermi level beyond 4.5$\%$ doping concentration ($x>0.045$). As a consequence, we can divide the phase diagram by $x=0.045$. When $x<0.045$, the VHS1 can be located at the Fermi level, and the CDW order may be formed. When $x>0.045$, all the energy of VHS1 is above the Fermi level, and the CDW order is hard to form. This result can well explain why no CDW anomaly occurs for all the Cs(V$_{1-x}$Ti$_x$)$_3$Sb$_5$ compounds with $x\geq 0.047$. As demonstrated by previous study~\cite{Tan2021}, the CDW instability can be identified by the imaginary frequency of the phonon spectrum. Here, we compare the calculated phonon spectrum under different doping level with $x=0$ and $x=0.07$, as shown in Figs.~\ref{fig:figure3}(g,h). For the undoped CsV$_3$Sb$_5$, unstable modes with imaginary frequencies exist at M and L. For the 7$\%$ doping concentration, these unstable modes disappear, which implies the absence of the CDW order. Combining all the results above, it can be concluded that the hole doping pushes the VHS1 above the Fermi level, and then lead to the disappearance of the CDW instabilities. Our results also reveal that the VHS1 plays a dominant role in the formation of the CDW order.

\begin{figure}[b]
  \centering
  \includegraphics[width=0.8\columnwidth]{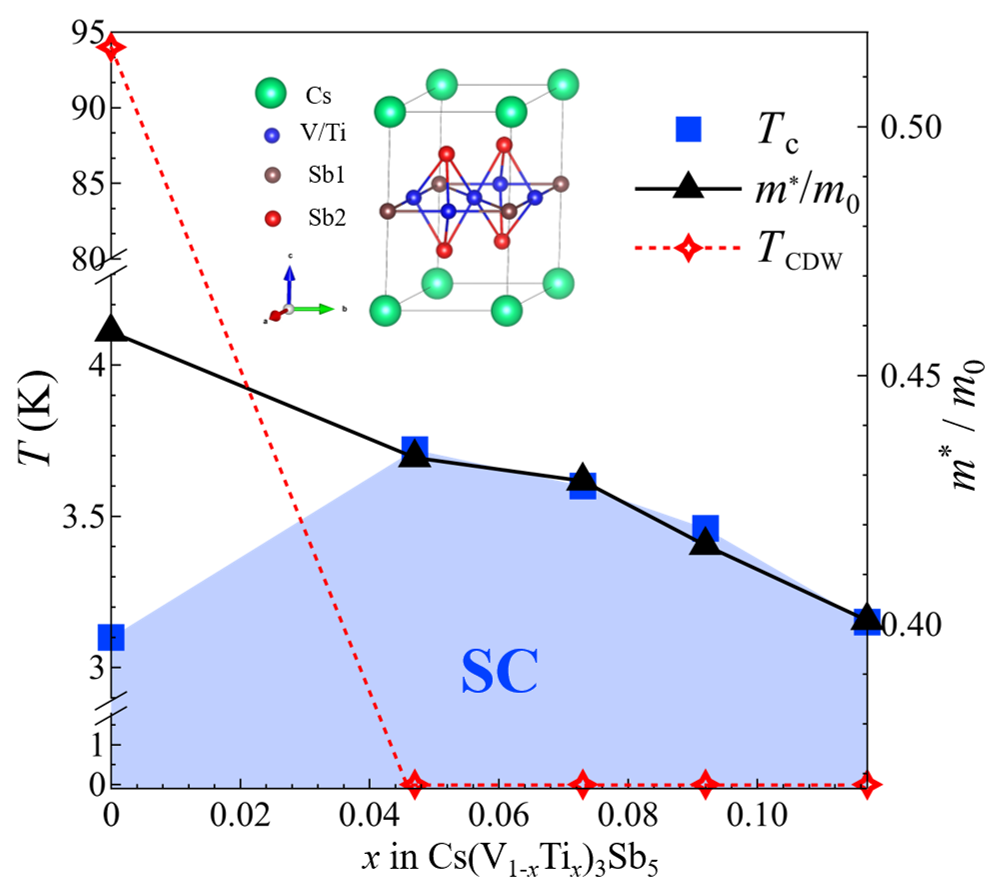}
  \caption{Doping evolution of the CDW
    order $T_{\rm cdw}$, superconducting transition temperatures $T_c$ and effective mass $m^*$ of
    the Sb-derived electron pockets at $\Gamma$ of
    Cs(V$_{1-x}$Ti$_x$)$_3$Sb$_5$.}
  \label{fig:figure4}
\end{figure}

The very different doping behaviors in Cs(V$_{1-x}$Ti$_x$)$_3$Sb$_5$ suggest
that the CDW and the superconducting orders in the parent compound have
different underlying electronic origins. Our results affirm the VHS saddle point
on the V bands leads to the CDW instability, consistent with the previous
theory~\cite{Tan2021}.  However, it is unlikely the reason for the superconductivity
since $T_c$ survives even when the doping of Ti pushes VHS1 above the Fermi
level, at odds with the previous speculation of the nature of pairing around the
VHS~\cite{Wu2021}. Recent NMR and magnetic penetration depth
measurements~\cite{Mu2021,Duan2021} suggest a conventional $s$-wave
superconducting paring in CsV$_3$Sb$_5$, which unlikely occurs on the V bands
regarding the robustness of the superconductivity upon doping in
Cs(V$_{1-x}$Ti$_x$)$_3$Sb$_5$. To investigate the relationship between the
superconductivity and the Sb electron pocket, we compare the effective mass
($m^*$) of the Sb electron pocket  at the $\Gamma$ point (C3 band in
Fig.~\ref{fig:figure2}) with the  superconducting transition temperatures $T_c$.
Due to the quasi-2D structure, $m^*$ is proportional to DOS,  and its doping
evolution agrees well with $T_c$ in Cs(V$_{1-x}$Ti$_x$)$_3$Sb$_5$ except the
parent compound, as shown  in Fig.~\ref{fig:figure4}. The similar doping
behaviors of $m^*$ and $T_c$ in the doped samples imply a BCS scenario,
consistent with the $s$-wave paring~\cite{Mu2021,Duan2021}. The parent compound
has the largest effective mass $m^*$, but with a suppressed $T_c$, consistent
with the competition relationship between the CDW order and the superconductivity~\cite{Wilson2021,Chen2021,Du2021,Yu2021,Chen2021b,Zhao2021}.

In conclusion, the doping evolution of the band structures in
Cs(V$_{1-x}$Ti$_x$)$_3$Sb$_5$ manifests that the van Hove singularity is the origin
of the CDW ordering, but not of the superconductivity which likely  occurs on
the Sb electron pocket. Although they have different origins, the charge order
and the superconductivity displays a competition relation, accounting for the
intertwined orders in the parent compoun CsV$_3$Sb$_5$.



\begin{acknowledgments}{This work is supported by National Natural Science Foundation of China (NSFC) (Grants Nos. 12074163 and 12004030), the Shenzhen High-level Special Fund (Grants No. G02206304 and No. G02206404), the Guangdong Innovative and Entrepreneurial Research Team Program (Grants No. 2017ZT07C062 and No. 2019ZT08C044), Shenzhen Science and Technology Program (Grant No. KQTD20190929173815000), the University Innovative Team in Guangdong Province (No. 2020KCXTD001), Shenzhen Key Laboratory of Advanced Quantum Functional Materials 
  and Devices (No. ZDSYS20190902092905285), Guangdong Basic and Applied Basic
  Research Foundation (No. 2020B1515120100), and China Postdoctoral Science
  Foundation (2020M682780).}
\end{acknowledgments}

\bibliography{VSC}

\begin{thebibliography}{43}%
\makeatletter
\providecommand \@ifxundefined [1]{%
 \@ifx{#1\undefined}
}%
\providecommand \@ifnum [1]{%
 \ifnum #1\expandafter \@firstoftwo
 \else \expandafter \@secondoftwo
 \fi
}%
\providecommand \@ifx [1]{%
 \ifx #1\expandafter \@firstoftwo
 \else \expandafter \@secondoftwo
 \fi
}%
\providecommand \natexlab [1]{#1}%
\providecommand \enquote  [1]{``#1''}%
\providecommand \bibnamefont  [1]{#1}%
\providecommand \bibfnamefont [1]{#1}%
\providecommand \citenamefont [1]{#1}%
\providecommand \href@noop [0]{\@secondoftwo}%
\providecommand \href [0]{\begingroup \@sanitize@url \@href}%
\providecommand \@href[1]{\@@startlink{#1}\@@href}%
\providecommand \@@href[1]{\endgroup#1\@@endlink}%
\providecommand \@sanitize@url [0]{\catcode `\\12\catcode `\$12\catcode
  `\&12\catcode `\#12\catcode `\^12\catcode `\_12\catcode `\%12\relax}%
\providecommand \@@startlink[1]{}%
\providecommand \@@endlink[0]{}%
\providecommand \url  [0]{\begingroup\@sanitize@url \@url }%
\providecommand \@url [1]{\endgroup\@href {#1}{\urlprefix }}%
\providecommand \urlprefix  [0]{URL }%
\providecommand \Eprint [0]{\href }%
\providecommand \doibase [0]{https://doi.org/}%
\providecommand \selectlanguage [0]{\@gobble}%
\providecommand \bibinfo  [0]{\@secondoftwo}%
\providecommand \bibfield  [0]{\@secondoftwo}%
\providecommand \translation [1]{[#1]}%
\providecommand \BibitemOpen [0]{}%
\providecommand \bibitemStop [0]{}%
\providecommand \bibitemNoStop [0]{.\EOS\space}%
\providecommand \EOS [0]{\spacefactor3000\relax}%
\providecommand \BibitemShut  [1]{\csname bibitem#1\endcsname}%
\let\auto@bib@innerbib\@empty
\bibitem [{\citenamefont {Isakov}\ \emph {et~al.}(2006)\citenamefont {Isakov},
  \citenamefont {Wessel}, \citenamefont {Melko}, \citenamefont {Sengupta},\
  and\ \citenamefont {Kim}}]{Isakov2006}%
  \BibitemOpen
  \bibfield  {author} {\bibinfo {author} {\bibfnamefont {S.~V.}\ \bibnamefont
  {Isakov}}, \bibinfo {author} {\bibfnamefont {S.}~\bibnamefont {Wessel}},
  \bibinfo {author} {\bibfnamefont {R.~G.}\ \bibnamefont {Melko}}, \bibinfo
  {author} {\bibfnamefont {K.}~\bibnamefont {Sengupta}},\ and\ \bibinfo
  {author} {\bibfnamefont {Y.~B.}\ \bibnamefont {Kim}},\ }\bibfield  {title}
  {\bibinfo {title} {Hard-core bosons on the kagome lattice: Valence-bond
  solids and their quantum melting},\ }\href
  {https://doi.org/10.1103/PhysRevLett.97.147202} {\bibfield  {journal}
  {\bibinfo  {journal} {Phys. Rev. Lett.}\ }\textbf {\bibinfo {volume} {97}},\
  \bibinfo {pages} {147202} (\bibinfo {year} {2006})}\BibitemShut {NoStop}%
\bibitem [{\citenamefont {Guo}\ and\ \citenamefont {Franz}(2009)}]{Guo2009}%
  \BibitemOpen
  \bibfield  {author} {\bibinfo {author} {\bibfnamefont {H.-M.}\ \bibnamefont
  {Guo}}\ and\ \bibinfo {author} {\bibfnamefont {M.}~\bibnamefont {Franz}},\
  }\bibfield  {title} {\bibinfo {title} {Topological insulator on the kagome
  lattice},\ }\href {https://doi.org/10.1103/PhysRevB.80.113102} {\bibfield
  {journal} {\bibinfo  {journal} {Phys. Rev. B}\ }\textbf {\bibinfo {volume}
  {80}},\ \bibinfo {pages} {113102} (\bibinfo {year} {2009})}\BibitemShut
  {NoStop}%
\bibitem [{\citenamefont {Tang}\ \emph {et~al.}(2011)\citenamefont {Tang},
  \citenamefont {Mei},\ and\ \citenamefont {Wen}}]{Tang2011}%
  \BibitemOpen
  \bibfield  {author} {\bibinfo {author} {\bibfnamefont {E.}~\bibnamefont
  {Tang}}, \bibinfo {author} {\bibfnamefont {J.-W.}\ \bibnamefont {Mei}},\ and\
  \bibinfo {author} {\bibfnamefont {X.-G.}\ \bibnamefont {Wen}},\ }\bibfield
  {title} {\bibinfo {title} {High-temperature fractional quantum hall states},\
  }\href {https://doi.org/10.1103/PhysRevLett.106.236802} {\bibfield  {journal}
  {\bibinfo  {journal} {Phys. Rev. Lett.}\ }\textbf {\bibinfo {volume} {106}},\
  \bibinfo {pages} {236802} (\bibinfo {year} {2011})}\BibitemShut {NoStop}%
\bibitem [{\citenamefont {Nakatsuji}\ \emph {et~al.}(2015)\citenamefont
  {Nakatsuji}, \citenamefont {Kiyohara},\ and\ \citenamefont
  {Higo}}]{Nakatsuji2015}%
  \BibitemOpen
  \bibfield  {author} {\bibinfo {author} {\bibfnamefont {S.}~\bibnamefont
  {Nakatsuji}}, \bibinfo {author} {\bibfnamefont {N.}~\bibnamefont
  {Kiyohara}},\ and\ \bibinfo {author} {\bibfnamefont {T.}~\bibnamefont
  {Higo}},\ }\bibfield  {title} {\bibinfo {title} {Large anomalous hall effect
  in a non-collinear antiferromagnet at room temperature},\ }\href
  {https://doi.org/10.1038/nature15723} {\bibfield  {journal} {\bibinfo
  {journal} {Nature}\ }\textbf {\bibinfo {volume} {527}},\ \bibinfo {pages}
  {212} (\bibinfo {year} {2015})}\BibitemShut {NoStop}%
\bibitem [{\citenamefont {Liu}\ \emph {et~al.}(2018)\citenamefont {Liu},
  \citenamefont {Sun}, \citenamefont {Kumar}, \citenamefont {Muechler},
  \citenamefont {Sun}, \citenamefont {Jiao}, \citenamefont {Yang},
  \citenamefont {Liu}, \citenamefont {Liang}, \citenamefont {Xu}, \citenamefont
  {Kroder}, \citenamefont {Süß}, \citenamefont {Borrmann}, \citenamefont
  {Shekhar}, \citenamefont {Wang}, \citenamefont {Xi}, \citenamefont {Wang},
  \citenamefont {Schnelle}, \citenamefont {Wirth}, \citenamefont {Chen},
  \citenamefont {Goennenwein},\ and\ \citenamefont {Felser}}]{Liu2018}%
  \BibitemOpen
  \bibfield  {author} {\bibinfo {author} {\bibfnamefont {E.}~\bibnamefont
  {Liu}}, \bibinfo {author} {\bibfnamefont {Y.}~\bibnamefont {Sun}}, \bibinfo
  {author} {\bibfnamefont {N.}~\bibnamefont {Kumar}}, \bibinfo {author}
  {\bibfnamefont {L.}~\bibnamefont {Muechler}}, \bibinfo {author}
  {\bibfnamefont {A.}~\bibnamefont {Sun}}, \bibinfo {author} {\bibfnamefont
  {L.}~\bibnamefont {Jiao}}, \bibinfo {author} {\bibfnamefont {S.-Y.}\
  \bibnamefont {Yang}}, \bibinfo {author} {\bibfnamefont {D.}~\bibnamefont
  {Liu}}, \bibinfo {author} {\bibfnamefont {A.}~\bibnamefont {Liang}}, \bibinfo
  {author} {\bibfnamefont {Q.}~\bibnamefont {Xu}}, \bibinfo {author}
  {\bibfnamefont {J.}~\bibnamefont {Kroder}}, \bibinfo {author} {\bibfnamefont
  {V.}~\bibnamefont {Süß}}, \bibinfo {author} {\bibfnamefont
  {H.}~\bibnamefont {Borrmann}}, \bibinfo {author} {\bibfnamefont
  {C.}~\bibnamefont {Shekhar}}, \bibinfo {author} {\bibfnamefont
  {Z.}~\bibnamefont {Wang}}, \bibinfo {author} {\bibfnamefont {C.}~\bibnamefont
  {Xi}}, \bibinfo {author} {\bibfnamefont {W.}~\bibnamefont {Wang}}, \bibinfo
  {author} {\bibfnamefont {W.}~\bibnamefont {Schnelle}}, \bibinfo {author}
  {\bibfnamefont {S.}~\bibnamefont {Wirth}}, \bibinfo {author} {\bibfnamefont
  {Y.}~\bibnamefont {Chen}}, \bibinfo {author} {\bibfnamefont {S.~T.~B.}\
  \bibnamefont {Goennenwein}},\ and\ \bibinfo {author} {\bibfnamefont
  {C.}~\bibnamefont {Felser}},\ }\bibfield  {title} {\bibinfo {title} {Giant
  anomalous hall effect in a ferromagnetic kagome-lattice semimetal},\ }\href
  {https://doi.org/10.1038/s41567-018-0234-5} {\bibfield  {journal} {\bibinfo
  {journal} {Nature Physics}\ }\textbf {\bibinfo {volume} {14}},\ \bibinfo
  {pages} {1125} (\bibinfo {year} {2018})}\BibitemShut {NoStop}%
\bibitem [{\citenamefont {Ye}\ \emph {et~al.}(2018)\citenamefont {Ye},
  \citenamefont {Kang}, \citenamefont {Liu}, \citenamefont {von Cube},
  \citenamefont {Wicker}, \citenamefont {Suzuki}, \citenamefont {Jozwiak},
  \citenamefont {Bostwick}, \citenamefont {Rotenberg}, \citenamefont {Bell},
  \citenamefont {Fu}, \citenamefont {Comin},\ and\ \citenamefont
  {Checkelsky}}]{Ye2018}%
  \BibitemOpen
  \bibfield  {author} {\bibinfo {author} {\bibfnamefont {L.}~\bibnamefont
  {Ye}}, \bibinfo {author} {\bibfnamefont {M.}~\bibnamefont {Kang}}, \bibinfo
  {author} {\bibfnamefont {J.}~\bibnamefont {Liu}}, \bibinfo {author}
  {\bibfnamefont {F.}~\bibnamefont {von Cube}}, \bibinfo {author}
  {\bibfnamefont {C.~R.}\ \bibnamefont {Wicker}}, \bibinfo {author}
  {\bibfnamefont {T.}~\bibnamefont {Suzuki}}, \bibinfo {author} {\bibfnamefont
  {C.}~\bibnamefont {Jozwiak}}, \bibinfo {author} {\bibfnamefont
  {A.}~\bibnamefont {Bostwick}}, \bibinfo {author} {\bibfnamefont
  {E.}~\bibnamefont {Rotenberg}}, \bibinfo {author} {\bibfnamefont {D.~C.}\
  \bibnamefont {Bell}}, \bibinfo {author} {\bibfnamefont {L.}~\bibnamefont
  {Fu}}, \bibinfo {author} {\bibfnamefont {R.}~\bibnamefont {Comin}},\ and\
  \bibinfo {author} {\bibfnamefont {J.~G.}\ \bibnamefont {Checkelsky}},\
  }\bibfield  {title} {\bibinfo {title} {Massive dirac fermions in a
  ferromagnetic kagome metal},\ }\href {https://doi.org/10.1038/nature25987}
  {\bibfield  {journal} {\bibinfo  {journal} {Nature}\ }\textbf {\bibinfo
  {volume} {555}},\ \bibinfo {pages} {638} (\bibinfo {year}
  {2018})}\BibitemShut {NoStop}%
\bibitem [{\citenamefont {Liu~D.}\ \emph {et~al.}(2019)\citenamefont {Liu~D.},
  \citenamefont {Liang~A.}, \citenamefont {Liu~E.}, \citenamefont {Xu~Q.},
  \citenamefont {Li~Y.}, \citenamefont {Chen}, \citenamefont {Pei},
  \citenamefont {Shi~W.}, \citenamefont {Mo~S.}, \citenamefont {Dudin},
  \citenamefont {Kim}, \citenamefont {Cacho}, \citenamefont {Li}, \citenamefont
  {Sun}, \citenamefont {Yang~L.}, \citenamefont {Liu~Z.}, \citenamefont {Parkin
  S.~S.}, \citenamefont {Felser},\ and\ \citenamefont {Chen~Y.}}]{LiuD.2019}%
  \BibitemOpen
  \bibfield  {author} {\bibinfo {author} {\bibfnamefont {F.}~\bibnamefont
  {Liu~D.}}, \bibinfo {author} {\bibfnamefont {J.}~\bibnamefont {Liang~A.}},
  \bibinfo {author} {\bibfnamefont {K.}~\bibnamefont {Liu~E.}}, \bibinfo
  {author} {\bibfnamefont {N.}~\bibnamefont {Xu~Q.}}, \bibinfo {author}
  {\bibfnamefont {W.}~\bibnamefont {Li~Y.}}, \bibinfo {author} {\bibfnamefont
  {C.}~\bibnamefont {Chen}}, \bibinfo {author} {\bibfnamefont {D.}~\bibnamefont
  {Pei}}, \bibinfo {author} {\bibfnamefont {J.}~\bibnamefont {Shi~W.}},
  \bibinfo {author} {\bibfnamefont {K.}~\bibnamefont {Mo~S.}}, \bibinfo
  {author} {\bibfnamefont {P.}~\bibnamefont {Dudin}}, \bibinfo {author}
  {\bibfnamefont {T.}~\bibnamefont {Kim}}, \bibinfo {author} {\bibfnamefont
  {C.}~\bibnamefont {Cacho}}, \bibinfo {author} {\bibfnamefont
  {G.}~\bibnamefont {Li}}, \bibinfo {author} {\bibfnamefont {Y.}~\bibnamefont
  {Sun}}, \bibinfo {author} {\bibfnamefont {X.}~\bibnamefont {Yang~L.}},
  \bibinfo {author} {\bibfnamefont {K.}~\bibnamefont {Liu~Z.}}, \bibinfo
  {author} {\bibfnamefont {P.}~\bibnamefont {Parkin S.~S.}}, \bibinfo {author}
  {\bibfnamefont {C.}~\bibnamefont {Felser}},\ and\ \bibinfo {author}
  {\bibfnamefont {L.}~\bibnamefont {Chen~Y.}},\ }\bibfield  {title} {\bibinfo
  {title} {Magnetic weyl semimetal phase in a kagomé crystal},\ }\href
  {https://doi.org/10.1126/science.aav2873} {\bibfield  {journal} {\bibinfo
  {journal} {Science}\ }\textbf {\bibinfo {volume} {365}},\ \bibinfo {pages}
  {1282} (\bibinfo {year} {2019})}\BibitemShut {NoStop}%
\bibitem [{\citenamefont {Yin}\ \emph {et~al.}(2020)\citenamefont {Yin},
  \citenamefont {Ma}, \citenamefont {Cochran}, \citenamefont {Xu},
  \citenamefont {Zhang}, \citenamefont {Tien}, \citenamefont {Shumiya},
  \citenamefont {Cheng}, \citenamefont {Jiang}, \citenamefont {Lian},
  \citenamefont {Song}, \citenamefont {Chang}, \citenamefont {Belopolski},
  \citenamefont {Multer}, \citenamefont {Litskevich}, \citenamefont {Cheng},
  \citenamefont {Yang}, \citenamefont {Swidler}, \citenamefont {Zhou},
  \citenamefont {Lin}, \citenamefont {Neupert}, \citenamefont {Wang},
  \citenamefont {Yao}, \citenamefont {Chang}, \citenamefont {Jia},\ and\
  \citenamefont {Zahid~Hasan}}]{Yin2020}%
  \BibitemOpen
  \bibfield  {author} {\bibinfo {author} {\bibfnamefont {J.-X.}\ \bibnamefont
  {Yin}}, \bibinfo {author} {\bibfnamefont {W.}~\bibnamefont {Ma}}, \bibinfo
  {author} {\bibfnamefont {T.~A.}\ \bibnamefont {Cochran}}, \bibinfo {author}
  {\bibfnamefont {X.}~\bibnamefont {Xu}}, \bibinfo {author} {\bibfnamefont
  {S.~S.}\ \bibnamefont {Zhang}}, \bibinfo {author} {\bibfnamefont {H.-J.}\
  \bibnamefont {Tien}}, \bibinfo {author} {\bibfnamefont {N.}~\bibnamefont
  {Shumiya}}, \bibinfo {author} {\bibfnamefont {G.}~\bibnamefont {Cheng}},
  \bibinfo {author} {\bibfnamefont {K.}~\bibnamefont {Jiang}}, \bibinfo
  {author} {\bibfnamefont {B.}~\bibnamefont {Lian}}, \bibinfo {author}
  {\bibfnamefont {Z.}~\bibnamefont {Song}}, \bibinfo {author} {\bibfnamefont
  {G.}~\bibnamefont {Chang}}, \bibinfo {author} {\bibfnamefont
  {I.}~\bibnamefont {Belopolski}}, \bibinfo {author} {\bibfnamefont
  {D.}~\bibnamefont {Multer}}, \bibinfo {author} {\bibfnamefont
  {M.}~\bibnamefont {Litskevich}}, \bibinfo {author} {\bibfnamefont {Z.-J.}\
  \bibnamefont {Cheng}}, \bibinfo {author} {\bibfnamefont {X.~P.}\ \bibnamefont
  {Yang}}, \bibinfo {author} {\bibfnamefont {B.}~\bibnamefont {Swidler}},
  \bibinfo {author} {\bibfnamefont {H.}~\bibnamefont {Zhou}}, \bibinfo {author}
  {\bibfnamefont {H.}~\bibnamefont {Lin}}, \bibinfo {author} {\bibfnamefont
  {T.}~\bibnamefont {Neupert}}, \bibinfo {author} {\bibfnamefont
  {Z.}~\bibnamefont {Wang}}, \bibinfo {author} {\bibfnamefont {N.}~\bibnamefont
  {Yao}}, \bibinfo {author} {\bibfnamefont {T.-R.}\ \bibnamefont {Chang}},
  \bibinfo {author} {\bibfnamefont {S.}~\bibnamefont {Jia}},\ and\ \bibinfo
  {author} {\bibfnamefont {M.}~\bibnamefont {Zahid~Hasan}},\ }\bibfield
  {title} {\bibinfo {title} {Quantum-limit chern topological magnetism in
  {TbMn$_6$Sn$_6$}},\ }\href {https://doi.org/10.1038/s41586-020-2482-7}
  {\bibfield  {journal} {\bibinfo  {journal} {Nature}\ }\textbf {\bibinfo
  {volume} {583}},\ \bibinfo {pages} {533} (\bibinfo {year}
  {2020})}\BibitemShut {NoStop}%
\bibitem [{\citenamefont {Ortiz}\ \emph {et~al.}(2019)\citenamefont {Ortiz},
  \citenamefont {Gomes}, \citenamefont {Morey}, \citenamefont {Winiarski},
  \citenamefont {Bordelon}, \citenamefont {Mangum}, \citenamefont {Oswald},
  \citenamefont {Rodriguez-Rivera}, \citenamefont {Neilson}, \citenamefont
  {Wilson}, \citenamefont {Ertekin}, \citenamefont {McQueen},\ and\
  \citenamefont {Toberer}}]{Ortiz2019}%
  \BibitemOpen
  \bibfield  {author} {\bibinfo {author} {\bibfnamefont {B.~R.}\ \bibnamefont
  {Ortiz}}, \bibinfo {author} {\bibfnamefont {L.~C.}\ \bibnamefont {Gomes}},
  \bibinfo {author} {\bibfnamefont {J.~R.}\ \bibnamefont {Morey}}, \bibinfo
  {author} {\bibfnamefont {M.}~\bibnamefont {Winiarski}}, \bibinfo {author}
  {\bibfnamefont {M.}~\bibnamefont {Bordelon}}, \bibinfo {author}
  {\bibfnamefont {J.~S.}\ \bibnamefont {Mangum}}, \bibinfo {author}
  {\bibfnamefont {I.~W.~H.}\ \bibnamefont {Oswald}}, \bibinfo {author}
  {\bibfnamefont {J.~A.}\ \bibnamefont {Rodriguez-Rivera}}, \bibinfo {author}
  {\bibfnamefont {J.~R.}\ \bibnamefont {Neilson}}, \bibinfo {author}
  {\bibfnamefont {S.~D.}\ \bibnamefont {Wilson}}, \bibinfo {author}
  {\bibfnamefont {E.}~\bibnamefont {Ertekin}}, \bibinfo {author} {\bibfnamefont
  {T.~M.}\ \bibnamefont {McQueen}},\ and\ \bibinfo {author} {\bibfnamefont
  {E.~S.}\ \bibnamefont {Toberer}},\ }\bibfield  {title} {\bibinfo {title} {New
  kagome prototype materials: discovery of
  {${\mathrm{KV}}_{3}{\mathrm{Sb}}_{5},{\mathrm{RbV}}_{3}{\mathrm{Sb}}_{5}$},
  and {${\mathrm{CsV}}_{3}{\mathrm{Sb}}_{5}$}},\ }\href
  {https://doi.org/10.1103/PhysRevMaterials.3.094407} {\bibfield  {journal}
  {\bibinfo  {journal} {Phys. Rev. Materials}\ }\textbf {\bibinfo {volume}
  {3}},\ \bibinfo {pages} {094407} (\bibinfo {year} {2019})}\BibitemShut
  {NoStop}%
\bibitem [{\citenamefont {Ortiz}\ \emph {et~al.}(2020)\citenamefont {Ortiz},
  \citenamefont {Teicher}, \citenamefont {Hu}, \citenamefont {Zuo},
  \citenamefont {Sarte}, \citenamefont {Schueller}, \citenamefont {Abeykoon},
  \citenamefont {Krogstad}, \citenamefont {Rosenkranz}, \citenamefont {Osborn},
  \citenamefont {Seshadri}, \citenamefont {Balents}, \citenamefont {He},\ and\
  \citenamefont {Wilson}}]{Ortiz2020}%
  \BibitemOpen
  \bibfield  {author} {\bibinfo {author} {\bibfnamefont {B.~R.}\ \bibnamefont
  {Ortiz}}, \bibinfo {author} {\bibfnamefont {S.~M.~L.}\ \bibnamefont
  {Teicher}}, \bibinfo {author} {\bibfnamefont {Y.}~\bibnamefont {Hu}},
  \bibinfo {author} {\bibfnamefont {J.~L.}\ \bibnamefont {Zuo}}, \bibinfo
  {author} {\bibfnamefont {P.~M.}\ \bibnamefont {Sarte}}, \bibinfo {author}
  {\bibfnamefont {E.~C.}\ \bibnamefont {Schueller}}, \bibinfo {author}
  {\bibfnamefont {A.~M.~M.}\ \bibnamefont {Abeykoon}}, \bibinfo {author}
  {\bibfnamefont {M.~J.}\ \bibnamefont {Krogstad}}, \bibinfo {author}
  {\bibfnamefont {S.}~\bibnamefont {Rosenkranz}}, \bibinfo {author}
  {\bibfnamefont {R.}~\bibnamefont {Osborn}}, \bibinfo {author} {\bibfnamefont
  {R.}~\bibnamefont {Seshadri}}, \bibinfo {author} {\bibfnamefont
  {L.}~\bibnamefont {Balents}}, \bibinfo {author} {\bibfnamefont
  {J.}~\bibnamefont {He}},\ and\ \bibinfo {author} {\bibfnamefont {S.~D.}\
  \bibnamefont {Wilson}},\ }\bibfield  {title} {\bibinfo {title}
  {{$\mathrm{Cs}{\mathrm{V}}_{3}{\mathrm{Sb}}_{5}$}: A {${\mathbb{Z}}_{2}$}
  topological kagome metal with a superconducting ground state},\ }\href
  {https://doi.org/10.1103/PhysRevLett.125.247002} {\bibfield  {journal}
  {\bibinfo  {journal} {Phys. Rev. Lett.}\ }\textbf {\bibinfo {volume} {125}},\
  \bibinfo {pages} {247002} (\bibinfo {year} {2020})}\BibitemShut {NoStop}%
\bibitem [{\citenamefont {Ortiz}\ \emph {et~al.}(2021)\citenamefont {Ortiz},
  \citenamefont {Sarte}, \citenamefont {Kenney}, \citenamefont {Graf},
  \citenamefont {Teicher}, \citenamefont {Seshadri},\ and\ \citenamefont
  {Wilson}}]{Ortiz2021}%
  \BibitemOpen
  \bibfield  {author} {\bibinfo {author} {\bibfnamefont {B.~R.}\ \bibnamefont
  {Ortiz}}, \bibinfo {author} {\bibfnamefont {P.~M.}\ \bibnamefont {Sarte}},
  \bibinfo {author} {\bibfnamefont {E.~M.}\ \bibnamefont {Kenney}}, \bibinfo
  {author} {\bibfnamefont {M.~J.}\ \bibnamefont {Graf}}, \bibinfo {author}
  {\bibfnamefont {S.~M.~L.}\ \bibnamefont {Teicher}}, \bibinfo {author}
  {\bibfnamefont {R.}~\bibnamefont {Seshadri}},\ and\ \bibinfo {author}
  {\bibfnamefont {S.~D.}\ \bibnamefont {Wilson}},\ }\bibfield  {title}
  {\bibinfo {title} {Superconductivity in the ${\mathbb{z}}_{2}$ kagome metal
  {${\mathrm{KV}}_{3}{\mathrm{Sb}}_{5}$}},\ }\href
  {https://doi.org/10.1103/PhysRevMaterials.5.034801} {\bibfield  {journal}
  {\bibinfo  {journal} {Phys. Rev. Materials}\ }\textbf {\bibinfo {volume}
  {5}},\ \bibinfo {pages} {034801} (\bibinfo {year} {2021})}\BibitemShut
  {NoStop}%
\bibitem [{\citenamefont {Johannes}\ and\ \citenamefont
  {Mazin}(2008)}]{Johannes2008}%
  \BibitemOpen
  \bibfield  {author} {\bibinfo {author} {\bibfnamefont {M.~D.}\ \bibnamefont
  {Johannes}}\ and\ \bibinfo {author} {\bibfnamefont {I.~I.}\ \bibnamefont
  {Mazin}},\ }\bibfield  {title} {\bibinfo {title} {Fermi surface nesting and
  the origin of charge density waves in metals},\ }\href
  {https://doi.org/10.1103/PhysRevB.77.165135} {\bibfield  {journal} {\bibinfo
  {journal} {Phys. Rev. B}\ }\textbf {\bibinfo {volume} {77}},\ \bibinfo
  {pages} {165135} (\bibinfo {year} {2008})}\BibitemShut {NoStop}%
\bibitem [{\citenamefont {Wang}\ \emph {et~al.}(2013)\citenamefont {Wang},
  \citenamefont {Li}, \citenamefont {Xiang},\ and\ \citenamefont
  {Wang}}]{Wang2013}%
  \BibitemOpen
  \bibfield  {author} {\bibinfo {author} {\bibfnamefont {W.-S.}\ \bibnamefont
  {Wang}}, \bibinfo {author} {\bibfnamefont {Z.-Z.}\ \bibnamefont {Li}},
  \bibinfo {author} {\bibfnamefont {Y.-Y.}\ \bibnamefont {Xiang}},\ and\
  \bibinfo {author} {\bibfnamefont {Q.-H.}\ \bibnamefont {Wang}},\ }\bibfield
  {title} {\bibinfo {title} {Competing electronic orders on kagome lattices at
  van hove filling},\ }\href {https://doi.org/10.1103/PhysRevB.87.115135}
  {\bibfield  {journal} {\bibinfo  {journal} {Phys. Rev. B}\ }\textbf {\bibinfo
  {volume} {87}},\ \bibinfo {pages} {115135} (\bibinfo {year}
  {2013})}\BibitemShut {NoStop}%
\bibitem [{\citenamefont {Kiesel}\ \emph {et~al.}(2013)\citenamefont {Kiesel},
  \citenamefont {Platt},\ and\ \citenamefont {Thomale}}]{Kiesel2013}%
  \BibitemOpen
  \bibfield  {author} {\bibinfo {author} {\bibfnamefont {M.~L.}\ \bibnamefont
  {Kiesel}}, \bibinfo {author} {\bibfnamefont {C.}~\bibnamefont {Platt}},\ and\
  \bibinfo {author} {\bibfnamefont {R.}~\bibnamefont {Thomale}},\ }\bibfield
  {title} {\bibinfo {title} {Unconventional fermi surface instabilities in the
  kagome hubbard model},\ }\href
  {https://doi.org/10.1103/PhysRevLett.110.126405} {\bibfield  {journal}
  {\bibinfo  {journal} {Phys. Rev. Lett.}\ }\textbf {\bibinfo {volume} {110}},\
  \bibinfo {pages} {126405} (\bibinfo {year} {2013})}\BibitemShut {NoStop}%
\bibitem [{\citenamefont {Chen}\ \emph
  {et~al.}(2021{\natexlab{a}})\citenamefont {Chen}, \citenamefont {Yang},
  \citenamefont {Hu}, \citenamefont {Zhao}, \citenamefont {Yuan}, \citenamefont
  {Xing}, \citenamefont {Qian}, \citenamefont {Huang}, \citenamefont {Li},
  \citenamefont {Ye}, \citenamefont {Ma}, \citenamefont {Ni}, \citenamefont
  {Zhang}, \citenamefont {Yin}, \citenamefont {Gong}, \citenamefont {Tu},
  \citenamefont {Lei}, \citenamefont {Tan}, \citenamefont {Zhou}, \citenamefont
  {Shen}, \citenamefont {Dong}, \citenamefont {Yan}, \citenamefont {Wang},\
  and\ \citenamefont {Gao}}]{Chen2021b}%
  \BibitemOpen
  \bibfield  {author} {\bibinfo {author} {\bibfnamefont {H.}~\bibnamefont
  {Chen}}, \bibinfo {author} {\bibfnamefont {H.}~\bibnamefont {Yang}}, \bibinfo
  {author} {\bibfnamefont {B.}~\bibnamefont {Hu}}, \bibinfo {author}
  {\bibfnamefont {Z.}~\bibnamefont {Zhao}}, \bibinfo {author} {\bibfnamefont
  {J.}~\bibnamefont {Yuan}}, \bibinfo {author} {\bibfnamefont {Y.}~\bibnamefont
  {Xing}}, \bibinfo {author} {\bibfnamefont {G.}~\bibnamefont {Qian}}, \bibinfo
  {author} {\bibfnamefont {Z.}~\bibnamefont {Huang}}, \bibinfo {author}
  {\bibfnamefont {G.}~\bibnamefont {Li}}, \bibinfo {author} {\bibfnamefont
  {Y.}~\bibnamefont {Ye}}, \bibinfo {author} {\bibfnamefont {S.}~\bibnamefont
  {Ma}}, \bibinfo {author} {\bibfnamefont {S.}~\bibnamefont {Ni}}, \bibinfo
  {author} {\bibfnamefont {H.}~\bibnamefont {Zhang}}, \bibinfo {author}
  {\bibfnamefont {Q.}~\bibnamefont {Yin}}, \bibinfo {author} {\bibfnamefont
  {C.}~\bibnamefont {Gong}}, \bibinfo {author} {\bibfnamefont {Z.}~\bibnamefont
  {Tu}}, \bibinfo {author} {\bibfnamefont {H.}~\bibnamefont {Lei}}, \bibinfo
  {author} {\bibfnamefont {H.}~\bibnamefont {Tan}}, \bibinfo {author}
  {\bibfnamefont {S.}~\bibnamefont {Zhou}}, \bibinfo {author} {\bibfnamefont
  {C.}~\bibnamefont {Shen}}, \bibinfo {author} {\bibfnamefont {X.}~\bibnamefont
  {Dong}}, \bibinfo {author} {\bibfnamefont {B.}~\bibnamefont {Yan}}, \bibinfo
  {author} {\bibfnamefont {Z.}~\bibnamefont {Wang}},\ and\ \bibinfo {author}
  {\bibfnamefont {H.-J.}\ \bibnamefont {Gao}},\ }\bibfield  {title} {\bibinfo
  {title} {Roton pair density wave in a strong-coupling kagome
  superconductor},\ }\bibfield  {journal} {\bibinfo  {journal} {Nature}\ }\href
  {https://doi.org/10.1038/s41586-021-03983-5} {10.1038/s41586-021-03983-5}
  (\bibinfo {year} {2021}{\natexlab{a}})\BibitemShut {NoStop}%
\bibitem [{\citenamefont {Zhao}\ \emph {et~al.}(2021)\citenamefont {Zhao},
  \citenamefont {Li}, \citenamefont {Ortiz}, \citenamefont {Teicher},
  \citenamefont {Park}, \citenamefont {Ye}, \citenamefont {Wang}, \citenamefont
  {Balents}, \citenamefont {Wilson},\ and\ \citenamefont
  {Zeljkovic}}]{Zhao2021}%
  \BibitemOpen
  \bibfield  {author} {\bibinfo {author} {\bibfnamefont {H.}~\bibnamefont
  {Zhao}}, \bibinfo {author} {\bibfnamefont {H.}~\bibnamefont {Li}}, \bibinfo
  {author} {\bibfnamefont {B.~R.}\ \bibnamefont {Ortiz}}, \bibinfo {author}
  {\bibfnamefont {S.~M.~L.}\ \bibnamefont {Teicher}}, \bibinfo {author}
  {\bibfnamefont {T.}~\bibnamefont {Park}}, \bibinfo {author} {\bibfnamefont
  {M.}~\bibnamefont {Ye}}, \bibinfo {author} {\bibfnamefont {Z.}~\bibnamefont
  {Wang}}, \bibinfo {author} {\bibfnamefont {L.}~\bibnamefont {Balents}},
  \bibinfo {author} {\bibfnamefont {S.~D.}\ \bibnamefont {Wilson}},\ and\
  \bibinfo {author} {\bibfnamefont {I.}~\bibnamefont {Zeljkovic}},\ }\bibfield
  {title} {\bibinfo {title} {Cascade of correlated electron states in a kagome
  superconductor {CsV$_3$Sb$_5$}},\ }\bibfield  {journal} {\bibinfo  {journal}
  {Nature}\ }\href {https://doi.org/10.1038/s41586-021-03946-w}
  {10.1038/s41586-021-03946-w} (\bibinfo {year} {2021})\BibitemShut {NoStop}%
\bibitem [{\citenamefont {Luo}\ \emph {et~al.}(2021{\natexlab{a}})\citenamefont
  {Luo}, \citenamefont {Gao}, \citenamefont {Liu}, \citenamefont {Gu},
  \citenamefont {Wu}, \citenamefont {Yi}, \citenamefont {Jia}, \citenamefont
  {Wu}, \citenamefont {Luo}, \citenamefont {Xu}, \citenamefont {Zhao},
  \citenamefont {Wang}, \citenamefont {Mao}, \citenamefont {Liu}, \citenamefont
  {Zhu}, \citenamefont {Shi}, \citenamefont {Jiang}, \citenamefont {Hu},
  \citenamefont {Xu},\ and\ \citenamefont {Zhou}}]{Luo2021}%
  \BibitemOpen
  \bibfield  {author} {\bibinfo {author} {\bibfnamefont {H.}~\bibnamefont
  {Luo}}, \bibinfo {author} {\bibfnamefont {Q.}~\bibnamefont {Gao}}, \bibinfo
  {author} {\bibfnamefont {H.}~\bibnamefont {Liu}}, \bibinfo {author}
  {\bibfnamefont {Y.}~\bibnamefont {Gu}}, \bibinfo {author} {\bibfnamefont
  {D.}~\bibnamefont {Wu}}, \bibinfo {author} {\bibfnamefont {C.}~\bibnamefont
  {Yi}}, \bibinfo {author} {\bibfnamefont {J.}~\bibnamefont {Jia}}, \bibinfo
  {author} {\bibfnamefont {S.}~\bibnamefont {Wu}}, \bibinfo {author}
  {\bibfnamefont {X.}~\bibnamefont {Luo}}, \bibinfo {author} {\bibfnamefont
  {Y.}~\bibnamefont {Xu}}, \bibinfo {author} {\bibfnamefont {L.}~\bibnamefont
  {Zhao}}, \bibinfo {author} {\bibfnamefont {Q.}~\bibnamefont {Wang}}, \bibinfo
  {author} {\bibfnamefont {H.}~\bibnamefont {Mao}}, \bibinfo {author}
  {\bibfnamefont {G.}~\bibnamefont {Liu}}, \bibinfo {author} {\bibfnamefont
  {Z.}~\bibnamefont {Zhu}}, \bibinfo {author} {\bibfnamefont {Y.}~\bibnamefont
  {Shi}}, \bibinfo {author} {\bibfnamefont {K.}~\bibnamefont {Jiang}}, \bibinfo
  {author} {\bibfnamefont {J.}~\bibnamefont {Hu}}, \bibinfo {author}
  {\bibfnamefont {Z.}~\bibnamefont {Xu}},\ and\ \bibinfo {author}
  {\bibfnamefont {X.~J.}\ \bibnamefont {Zhou}},\ }\href@noop {} {\bibinfo
  {title} {Electronic nature of charge density wave and electron-phonon
  coupling in kagome superconductor {KV$_3$Sb$_5$}}} (\bibinfo {year}
  {2021}{\natexlab{a}}),\ \Eprint {https://arxiv.org/abs/2107.02688}
  {arXiv:2107.02688 [cond-mat.supr-con]} \BibitemShut {NoStop}%
\bibitem [{\citenamefont {Li}\ \emph {et~al.}(2021)\citenamefont {Li},
  \citenamefont {Zhang}, \citenamefont {Yilmaz}, \citenamefont {Pai},
  \citenamefont {Marvinney}, \citenamefont {Said}, \citenamefont {Yin},
  \citenamefont {Gong}, \citenamefont {Tu}, \citenamefont {Vescovo},
  \citenamefont {Nelson}, \citenamefont {Moore}, \citenamefont {Murakami},
  \citenamefont {Lei}, \citenamefont {Lee}, \citenamefont {Lawrie},\ and\
  \citenamefont {Miao}}]{Li2021}%
  \BibitemOpen
  \bibfield  {author} {\bibinfo {author} {\bibfnamefont {H.}~\bibnamefont
  {Li}}, \bibinfo {author} {\bibfnamefont {T.~T.}\ \bibnamefont {Zhang}},
  \bibinfo {author} {\bibfnamefont {T.}~\bibnamefont {Yilmaz}}, \bibinfo
  {author} {\bibfnamefont {Y.~Y.}\ \bibnamefont {Pai}}, \bibinfo {author}
  {\bibfnamefont {C.~E.}\ \bibnamefont {Marvinney}}, \bibinfo {author}
  {\bibfnamefont {A.}~\bibnamefont {Said}}, \bibinfo {author} {\bibfnamefont
  {Q.~W.}\ \bibnamefont {Yin}}, \bibinfo {author} {\bibfnamefont {C.~S.}\
  \bibnamefont {Gong}}, \bibinfo {author} {\bibfnamefont {Z.~J.}\ \bibnamefont
  {Tu}}, \bibinfo {author} {\bibfnamefont {E.}~\bibnamefont {Vescovo}},
  \bibinfo {author} {\bibfnamefont {C.~S.}\ \bibnamefont {Nelson}}, \bibinfo
  {author} {\bibfnamefont {R.~G.}\ \bibnamefont {Moore}}, \bibinfo {author}
  {\bibfnamefont {S.}~\bibnamefont {Murakami}}, \bibinfo {author}
  {\bibfnamefont {H.~C.}\ \bibnamefont {Lei}}, \bibinfo {author} {\bibfnamefont
  {H.~N.}\ \bibnamefont {Lee}}, \bibinfo {author} {\bibfnamefont {B.~J.}\
  \bibnamefont {Lawrie}},\ and\ \bibinfo {author} {\bibfnamefont
  {H.}~\bibnamefont {Miao}},\ }\bibfield  {title} {\bibinfo {title}
  {Observation of unconventional charge density wave without acoustic phonon
  anomaly in kagome superconductors {${A\mathrm{V}}_{3}{\mathrm{Sb}}_{5}$
  ($A=\mathrm{Rb}$, Cs)}},\ }\href {https://doi.org/10.1103/PhysRevX.11.031050}
  {\bibfield  {journal} {\bibinfo  {journal} {Phys. Rev. X}\ }\textbf {\bibinfo
  {volume} {11}},\ \bibinfo {pages} {031050} (\bibinfo {year}
  {2021})}\BibitemShut {NoStop}%
\bibitem [{\citenamefont {Kang}\ \emph {et~al.}(2021)\citenamefont {Kang},
  \citenamefont {Fang}, \citenamefont {Kim}, \citenamefont {Ortiz},
  \citenamefont {Yoo}, \citenamefont {Park}, \citenamefont {Wilson},
  \citenamefont {Park},\ and\ \citenamefont {Comin}}]{Kang2021}%
  \BibitemOpen
  \bibfield  {author} {\bibinfo {author} {\bibfnamefont {M.}~\bibnamefont
  {Kang}}, \bibinfo {author} {\bibfnamefont {S.}~\bibnamefont {Fang}}, \bibinfo
  {author} {\bibfnamefont {J.-K.}\ \bibnamefont {Kim}}, \bibinfo {author}
  {\bibfnamefont {B.~R.}\ \bibnamefont {Ortiz}}, \bibinfo {author}
  {\bibfnamefont {J.}~\bibnamefont {Yoo}}, \bibinfo {author} {\bibfnamefont
  {B.-G.}\ \bibnamefont {Park}}, \bibinfo {author} {\bibfnamefont {S.~D.}\
  \bibnamefont {Wilson}}, \bibinfo {author} {\bibfnamefont {J.-H.}\
  \bibnamefont {Park}},\ and\ \bibinfo {author} {\bibfnamefont
  {R.}~\bibnamefont {Comin}},\ }\href@noop {} {\bibinfo {title} {Twofold van
  hove singularity and origin of charge order in topological kagome
  superconductor {CsV$_3$Sb$_5$}}} (\bibinfo {year} {2021}),\ \Eprint
  {https://arxiv.org/abs/2105.01689} {arXiv:2105.01689 [cond-mat.str-el]}
  \BibitemShut {NoStop}%
\bibitem [{\citenamefont {Lou}\ \emph {et~al.}(2021)\citenamefont {Lou},
  \citenamefont {Fedorov}, \citenamefont {Yin}, \citenamefont {Kuibarov},
  \citenamefont {Tu}, \citenamefont {Gong}, \citenamefont {Schwier},
  \citenamefont {Büchner}, \citenamefont {Lei},\ and\ \citenamefont
  {Borisenko}}]{Lou2021}%
  \BibitemOpen
  \bibfield  {author} {\bibinfo {author} {\bibfnamefont {R.}~\bibnamefont
  {Lou}}, \bibinfo {author} {\bibfnamefont {A.}~\bibnamefont {Fedorov}},
  \bibinfo {author} {\bibfnamefont {Q.}~\bibnamefont {Yin}}, \bibinfo {author}
  {\bibfnamefont {A.}~\bibnamefont {Kuibarov}}, \bibinfo {author}
  {\bibfnamefont {Z.}~\bibnamefont {Tu}}, \bibinfo {author} {\bibfnamefont
  {C.}~\bibnamefont {Gong}}, \bibinfo {author} {\bibfnamefont {E.~F.}\
  \bibnamefont {Schwier}}, \bibinfo {author} {\bibfnamefont {B.}~\bibnamefont
  {Büchner}}, \bibinfo {author} {\bibfnamefont {H.}~\bibnamefont {Lei}},\ and\
  \bibinfo {author} {\bibfnamefont {S.}~\bibnamefont {Borisenko}},\ }\href@noop
  {} {\bibinfo {title} {Charge-density-wave-induced peak-dip-hump structure and
  flat band in the kagome superconductor {CsV$_{3}$Sb$_{5}$}}} (\bibinfo {year}
  {2021}),\ \Eprint {https://arxiv.org/abs/2106.06497} {arXiv:2106.06497
  [cond-mat.supr-con]} \BibitemShut {NoStop}%
\bibitem [{\citenamefont {Cho}\ \emph {et~al.}(2021)\citenamefont {Cho},
  \citenamefont {Ma}, \citenamefont {Xia}, \citenamefont {Yang}, \citenamefont
  {Liu}, \citenamefont {Huang}, \citenamefont {Jiang}, \citenamefont {Lu},
  \citenamefont {Liu}, \citenamefont {Liu}, \citenamefont {Jia}, \citenamefont
  {Guo}, \citenamefont {Liu},\ and\ \citenamefont {Shen}}]{Cho2021}%
  \BibitemOpen
  \bibfield  {author} {\bibinfo {author} {\bibfnamefont {S.}~\bibnamefont
  {Cho}}, \bibinfo {author} {\bibfnamefont {H.}~\bibnamefont {Ma}}, \bibinfo
  {author} {\bibfnamefont {W.}~\bibnamefont {Xia}}, \bibinfo {author}
  {\bibfnamefont {Y.}~\bibnamefont {Yang}}, \bibinfo {author} {\bibfnamefont
  {Z.}~\bibnamefont {Liu}}, \bibinfo {author} {\bibfnamefont {Z.}~\bibnamefont
  {Huang}}, \bibinfo {author} {\bibfnamefont {Z.}~\bibnamefont {Jiang}},
  \bibinfo {author} {\bibfnamefont {X.}~\bibnamefont {Lu}}, \bibinfo {author}
  {\bibfnamefont {J.}~\bibnamefont {Liu}}, \bibinfo {author} {\bibfnamefont
  {Z.}~\bibnamefont {Liu}}, \bibinfo {author} {\bibfnamefont {J.}~\bibnamefont
  {Jia}}, \bibinfo {author} {\bibfnamefont {Y.}~\bibnamefont {Guo}}, \bibinfo
  {author} {\bibfnamefont {J.}~\bibnamefont {Liu}},\ and\ \bibinfo {author}
  {\bibfnamefont {D.}~\bibnamefont {Shen}},\ }\href@noop {} {\bibinfo {title}
  {Emergence of new van hove singularities in the charge density wave state of
  a topological kagome metal {RbV$_3$Sb$_5$}}} (\bibinfo {year} {2021}),\
  \Eprint {https://arxiv.org/abs/2105.05117} {arXiv:2105.05117
  [cond-mat.supr-con]} \BibitemShut {NoStop}%
\bibitem [{\citenamefont {Hu}\ \emph {et~al.}(2021{\natexlab{a}})\citenamefont
  {Hu}, \citenamefont {Wu}, \citenamefont {Ortiz}, \citenamefont {Ju},
  \citenamefont {Han}, \citenamefont {Ma}, \citenamefont {Plumb}, \citenamefont
  {Radovic}, \citenamefont {Thomale}, \citenamefont {Wilson}, \citenamefont
  {Schnyder},\ and\ \citenamefont {Shi}}]{Hu2021}%
  \BibitemOpen
  \bibfield  {author} {\bibinfo {author} {\bibfnamefont {Y.}~\bibnamefont
  {Hu}}, \bibinfo {author} {\bibfnamefont {X.}~\bibnamefont {Wu}}, \bibinfo
  {author} {\bibfnamefont {B.~R.}\ \bibnamefont {Ortiz}}, \bibinfo {author}
  {\bibfnamefont {S.}~\bibnamefont {Ju}}, \bibinfo {author} {\bibfnamefont
  {X.}~\bibnamefont {Han}}, \bibinfo {author} {\bibfnamefont {J.~Z.}\
  \bibnamefont {Ma}}, \bibinfo {author} {\bibfnamefont {N.~C.}\ \bibnamefont
  {Plumb}}, \bibinfo {author} {\bibfnamefont {M.}~\bibnamefont {Radovic}},
  \bibinfo {author} {\bibfnamefont {R.}~\bibnamefont {Thomale}}, \bibinfo
  {author} {\bibfnamefont {S.~D.}\ \bibnamefont {Wilson}}, \bibinfo {author}
  {\bibfnamefont {A.~P.}\ \bibnamefont {Schnyder}},\ and\ \bibinfo {author}
  {\bibfnamefont {M.}~\bibnamefont {Shi}},\ }\href@noop {} {\bibinfo {title}
  {Rich nature of van hove singularities in kagome superconductor
  {CsV$_3$Sb$_5$}}} (\bibinfo {year} {2021}{\natexlab{a}}),\ \Eprint
  {https://arxiv.org/abs/2106.05922} {arXiv:2106.05922 [cond-mat.supr-con]}
  \BibitemShut {NoStop}%
\bibitem [{\citenamefont {Hu}\ \emph {et~al.}(2021{\natexlab{b}})\citenamefont
  {Hu}, \citenamefont {Teicher}, \citenamefont {Ortiz}, \citenamefont {Luo},
  \citenamefont {Peng}, \citenamefont {Huai}, \citenamefont {Ma}, \citenamefont
  {Plumb}, \citenamefont {Wilson}, \citenamefont {He},\ and\ \citenamefont
  {Shi}}]{Hu2021a}%
  \BibitemOpen
  \bibfield  {author} {\bibinfo {author} {\bibfnamefont {Y.}~\bibnamefont
  {Hu}}, \bibinfo {author} {\bibfnamefont {S.~M.~L.}\ \bibnamefont {Teicher}},
  \bibinfo {author} {\bibfnamefont {B.~R.}\ \bibnamefont {Ortiz}}, \bibinfo
  {author} {\bibfnamefont {Y.}~\bibnamefont {Luo}}, \bibinfo {author}
  {\bibfnamefont {S.}~\bibnamefont {Peng}}, \bibinfo {author} {\bibfnamefont
  {L.}~\bibnamefont {Huai}}, \bibinfo {author} {\bibfnamefont {J.~Z.}\
  \bibnamefont {Ma}}, \bibinfo {author} {\bibfnamefont {N.~C.}\ \bibnamefont
  {Plumb}}, \bibinfo {author} {\bibfnamefont {S.~D.}\ \bibnamefont {Wilson}},
  \bibinfo {author} {\bibfnamefont {J.~F.}\ \bibnamefont {He}},\ and\ \bibinfo
  {author} {\bibfnamefont {M.}~\bibnamefont {Shi}},\ }\href@noop {} {\bibinfo
  {title} {Charge-order-assisted topological surface states and flat bands in
  the kagome superconductor {CsV$_3$Sb$_5$}}} (\bibinfo {year}
  {2021}{\natexlab{b}}),\ \Eprint {https://arxiv.org/abs/2104.12725}
  {arXiv:2104.12725 [cond-mat.supr-con]} \BibitemShut {NoStop}%
\bibitem [{\citenamefont {Cai}\ \emph {et~al.}(2021)\citenamefont {Cai},
  \citenamefont {Wang}, \citenamefont {Hao}, \citenamefont {Liu}, \citenamefont
  {Ma}, \citenamefont {Shen}, \citenamefont {Jiang}, \citenamefont {Yang},
  \citenamefont {Liu}, \citenamefont {Jiang}, \citenamefont {Liu},
  \citenamefont {Ye}, \citenamefont {Shen}, \citenamefont {Sun}, \citenamefont
  {Chen}, \citenamefont {Wang}, \citenamefont {Liu}, \citenamefont {Lin},
  \citenamefont {Wang}, \citenamefont {Huang}, \citenamefont {Mei},\ and\
  \citenamefont {Chen}}]{Cai2021}%
  \BibitemOpen
  \bibfield  {author} {\bibinfo {author} {\bibfnamefont {Y.}~\bibnamefont
  {Cai}}, \bibinfo {author} {\bibfnamefont {Y.}~\bibnamefont {Wang}}, \bibinfo
  {author} {\bibfnamefont {Z.}~\bibnamefont {Hao}}, \bibinfo {author}
  {\bibfnamefont {Y.}~\bibnamefont {Liu}}, \bibinfo {author} {\bibfnamefont
  {X.-M.}\ \bibnamefont {Ma}}, \bibinfo {author} {\bibfnamefont
  {Z.}~\bibnamefont {Shen}}, \bibinfo {author} {\bibfnamefont {Z.}~\bibnamefont
  {Jiang}}, \bibinfo {author} {\bibfnamefont {Y.}~\bibnamefont {Yang}},
  \bibinfo {author} {\bibfnamefont {W.}~\bibnamefont {Liu}}, \bibinfo {author}
  {\bibfnamefont {Q.}~\bibnamefont {Jiang}}, \bibinfo {author} {\bibfnamefont
  {Z.}~\bibnamefont {Liu}}, \bibinfo {author} {\bibfnamefont {M.}~\bibnamefont
  {Ye}}, \bibinfo {author} {\bibfnamefont {D.}~\bibnamefont {Shen}}, \bibinfo
  {author} {\bibfnamefont {Z.}~\bibnamefont {Sun}}, \bibinfo {author}
  {\bibfnamefont {J.}~\bibnamefont {Chen}}, \bibinfo {author} {\bibfnamefont
  {L.}~\bibnamefont {Wang}}, \bibinfo {author} {\bibfnamefont {C.}~\bibnamefont
  {Liu}}, \bibinfo {author} {\bibfnamefont {J.}~\bibnamefont {Lin}}, \bibinfo
  {author} {\bibfnamefont {J.}~\bibnamefont {Wang}}, \bibinfo {author}
  {\bibfnamefont {B.}~\bibnamefont {Huang}}, \bibinfo {author} {\bibfnamefont
  {J.-W.}\ \bibnamefont {Mei}},\ and\ \bibinfo {author} {\bibfnamefont
  {C.}~\bibnamefont {Chen}},\ }\href@noop {} {\bibinfo {title} {Emergence of
  quantum confinement in topological kagome superconductor {CsV$_3$Sb$_5$}
  family}} (\bibinfo {year} {2021}),\ \Eprint
  {https://arxiv.org/abs/2109.12778} {arXiv:2109.12778 [cond-mat.str-el]}
  \BibitemShut {NoStop}%
\bibitem [{\citenamefont {Luo}\ \emph {et~al.}(2021{\natexlab{b}})\citenamefont
  {Luo}, \citenamefont {Peng}, \citenamefont {Teicher}, \citenamefont {Huai},
  \citenamefont {Hu}, \citenamefont {Ortiz}, \citenamefont {Wei}, \citenamefont
  {Shen}, \citenamefont {Ou}, \citenamefont {Wang}, \citenamefont {Miao},
  \citenamefont {Guo}, \citenamefont {Shi}, \citenamefont {Wilson},\ and\
  \citenamefont {He}}]{Luo2021a}%
  \BibitemOpen
  \bibfield  {author} {\bibinfo {author} {\bibfnamefont {Y.}~\bibnamefont
  {Luo}}, \bibinfo {author} {\bibfnamefont {S.}~\bibnamefont {Peng}}, \bibinfo
  {author} {\bibfnamefont {S.~M.~L.}\ \bibnamefont {Teicher}}, \bibinfo
  {author} {\bibfnamefont {L.}~\bibnamefont {Huai}}, \bibinfo {author}
  {\bibfnamefont {Y.}~\bibnamefont {Hu}}, \bibinfo {author} {\bibfnamefont
  {B.~R.}\ \bibnamefont {Ortiz}}, \bibinfo {author} {\bibfnamefont
  {Z.}~\bibnamefont {Wei}}, \bibinfo {author} {\bibfnamefont {J.}~\bibnamefont
  {Shen}}, \bibinfo {author} {\bibfnamefont {Z.}~\bibnamefont {Ou}}, \bibinfo
  {author} {\bibfnamefont {B.}~\bibnamefont {Wang}}, \bibinfo {author}
  {\bibfnamefont {Y.}~\bibnamefont {Miao}}, \bibinfo {author} {\bibfnamefont
  {M.}~\bibnamefont {Guo}}, \bibinfo {author} {\bibfnamefont {M.}~\bibnamefont
  {Shi}}, \bibinfo {author} {\bibfnamefont {S.~D.}\ \bibnamefont {Wilson}},\
  and\ \bibinfo {author} {\bibfnamefont {J.~F.}\ \bibnamefont {He}},\
  }\href@noop {} {\bibinfo {title} {Distinct band reconstructions in kagome
  superconductor {CsV$_3$Sb$_5$}}} (\bibinfo {year} {2021}{\natexlab{b}}),\
  \Eprint {https://arxiv.org/abs/2106.01248} {arXiv:2106.01248
  [cond-mat.str-el]} \BibitemShut {NoStop}%
\bibitem [{\citenamefont {Liang}\ \emph {et~al.}(2021)\citenamefont {Liang},
  \citenamefont {Hou}, \citenamefont {Zhang}, \citenamefont {Ma}, \citenamefont
  {Wu}, \citenamefont {Zhang}, \citenamefont {Yu}, \citenamefont {Ying},
  \citenamefont {Jiang}, \citenamefont {Shan}, \citenamefont {Wang},\ and\
  \citenamefont {Chen}}]{Liang2021}%
  \BibitemOpen
  \bibfield  {author} {\bibinfo {author} {\bibfnamefont {Z.}~\bibnamefont
  {Liang}}, \bibinfo {author} {\bibfnamefont {X.}~\bibnamefont {Hou}}, \bibinfo
  {author} {\bibfnamefont {F.}~\bibnamefont {Zhang}}, \bibinfo {author}
  {\bibfnamefont {W.}~\bibnamefont {Ma}}, \bibinfo {author} {\bibfnamefont
  {P.}~\bibnamefont {Wu}}, \bibinfo {author} {\bibfnamefont {Z.}~\bibnamefont
  {Zhang}}, \bibinfo {author} {\bibfnamefont {F.}~\bibnamefont {Yu}}, \bibinfo
  {author} {\bibfnamefont {J.-J.}\ \bibnamefont {Ying}}, \bibinfo {author}
  {\bibfnamefont {K.}~\bibnamefont {Jiang}}, \bibinfo {author} {\bibfnamefont
  {L.}~\bibnamefont {Shan}}, \bibinfo {author} {\bibfnamefont {Z.}~\bibnamefont
  {Wang}},\ and\ \bibinfo {author} {\bibfnamefont {X.-H.}\ \bibnamefont
  {Chen}},\ }\bibfield  {title} {\bibinfo {title} {Three-dimensional charge
  density wave and surface-dependent vortex-core states in a kagome
  superconductor {${\mathrm{CsV}}_{3}{\mathrm{Sb}}_{5}$}},\ }\href
  {https://doi.org/10.1103/PhysRevX.11.031026} {\bibfield  {journal} {\bibinfo
  {journal} {Phys. Rev. X}\ }\textbf {\bibinfo {volume} {11}},\ \bibinfo
  {pages} {031026} (\bibinfo {year} {2021})}\BibitemShut {NoStop}%
\bibitem [{\citenamefont {Jiang}\ \emph {et~al.}(2021)\citenamefont {Jiang},
  \citenamefont {Yin}, \citenamefont {Denner}, \citenamefont {Shumiya},
  \citenamefont {Ortiz}, \citenamefont {Xu}, \citenamefont {Guguchia},
  \citenamefont {He}, \citenamefont {Hossain}, \citenamefont {Liu},
  \citenamefont {Ruff}, \citenamefont {Kautzsch}, \citenamefont {Zhang},
  \citenamefont {Chang}, \citenamefont {Belopolski}, \citenamefont {Zhang},
  \citenamefont {Cochran}, \citenamefont {Multer}, \citenamefont {Litskevich},
  \citenamefont {Cheng}, \citenamefont {Yang}, \citenamefont {Wang},
  \citenamefont {Thomale}, \citenamefont {Neupert}, \citenamefont {Wilson},\
  and\ \citenamefont {Hasan}}]{Jiang2021}%
  \BibitemOpen
  \bibfield  {author} {\bibinfo {author} {\bibfnamefont {Y.-X.}\ \bibnamefont
  {Jiang}}, \bibinfo {author} {\bibfnamefont {J.-X.}\ \bibnamefont {Yin}},
  \bibinfo {author} {\bibfnamefont {M.~M.}\ \bibnamefont {Denner}}, \bibinfo
  {author} {\bibfnamefont {N.}~\bibnamefont {Shumiya}}, \bibinfo {author}
  {\bibfnamefont {B.~R.}\ \bibnamefont {Ortiz}}, \bibinfo {author}
  {\bibfnamefont {G.}~\bibnamefont {Xu}}, \bibinfo {author} {\bibfnamefont
  {Z.}~\bibnamefont {Guguchia}}, \bibinfo {author} {\bibfnamefont
  {J.}~\bibnamefont {He}}, \bibinfo {author} {\bibfnamefont {M.~S.}\
  \bibnamefont {Hossain}}, \bibinfo {author} {\bibfnamefont {X.}~\bibnamefont
  {Liu}}, \bibinfo {author} {\bibfnamefont {J.}~\bibnamefont {Ruff}}, \bibinfo
  {author} {\bibfnamefont {L.}~\bibnamefont {Kautzsch}}, \bibinfo {author}
  {\bibfnamefont {S.~S.}\ \bibnamefont {Zhang}}, \bibinfo {author}
  {\bibfnamefont {G.}~\bibnamefont {Chang}}, \bibinfo {author} {\bibfnamefont
  {I.}~\bibnamefont {Belopolski}}, \bibinfo {author} {\bibfnamefont
  {Q.}~\bibnamefont {Zhang}}, \bibinfo {author} {\bibfnamefont {T.~A.}\
  \bibnamefont {Cochran}}, \bibinfo {author} {\bibfnamefont {D.}~\bibnamefont
  {Multer}}, \bibinfo {author} {\bibfnamefont {M.}~\bibnamefont {Litskevich}},
  \bibinfo {author} {\bibfnamefont {Z.-J.}\ \bibnamefont {Cheng}}, \bibinfo
  {author} {\bibfnamefont {X.~P.}\ \bibnamefont {Yang}}, \bibinfo {author}
  {\bibfnamefont {Z.}~\bibnamefont {Wang}}, \bibinfo {author} {\bibfnamefont
  {R.}~\bibnamefont {Thomale}}, \bibinfo {author} {\bibfnamefont
  {T.}~\bibnamefont {Neupert}}, \bibinfo {author} {\bibfnamefont {S.~D.}\
  \bibnamefont {Wilson}},\ and\ \bibinfo {author} {\bibfnamefont {M.~Z.}\
  \bibnamefont {Hasan}},\ }\bibfield  {title} {\bibinfo {title} {Unconventional
  chiral charge order in kagome superconductor {KV$_3$Sb$_5$}},\ }\href
  {https://doi.org/10.1038/s41563-021-01034-y} {\bibfield  {journal} {\bibinfo
  {journal} {Nature Materials}\ }\textbf {\bibinfo {volume} {20}},\ \bibinfo
  {pages} {1353} (\bibinfo {year} {2021})}\BibitemShut {NoStop}%
\bibitem [{\citenamefont {Feng}\ \emph {et~al.}(2021)\citenamefont {Feng},
  \citenamefont {Jiang}, \citenamefont {Wang},\ and\ \citenamefont
  {Hu}}]{Feng2021}%
  \BibitemOpen
  \bibfield  {author} {\bibinfo {author} {\bibfnamefont {X.}~\bibnamefont
  {Feng}}, \bibinfo {author} {\bibfnamefont {K.}~\bibnamefont {Jiang}},
  \bibinfo {author} {\bibfnamefont {Z.}~\bibnamefont {Wang}},\ and\ \bibinfo
  {author} {\bibfnamefont {J.}~\bibnamefont {Hu}},\ }\bibfield  {title}
  {\bibinfo {title} {Chiral flux phase in the kagome superconductor
  {$A$V$_3$Sb$_5$}},\ }\href
  {https://doi.org/https://doi.org/10.1016/j.scib.2021.04.043} {\bibfield
  {journal} {\bibinfo  {journal} {Science Bulletin}\ }\textbf {\bibinfo
  {volume} {66}},\ \bibinfo {pages} {1384} (\bibinfo {year}
  {2021})}\BibitemShut {NoStop}%
\bibitem [{\citenamefont {Denner}\ \emph {et~al.}(2021)\citenamefont {Denner},
  \citenamefont {Thomale},\ and\ \citenamefont {Neupert}}]{Denner2021}%
  \BibitemOpen
  \bibfield  {author} {\bibinfo {author} {\bibfnamefont {M.~M.}\ \bibnamefont
  {Denner}}, \bibinfo {author} {\bibfnamefont {R.}~\bibnamefont {Thomale}},\
  and\ \bibinfo {author} {\bibfnamefont {T.}~\bibnamefont {Neupert}},\
  }\href@noop {} {\bibinfo {title} {Analysis of charge order in the kagome
  metal {$A$V$_3$Sb$_5$ ($A=$ K,Rb,Cs)}}} (\bibinfo {year} {2021}),\ \Eprint
  {https://arxiv.org/abs/2103.14045} {arXiv:2103.14045 [cond-mat.str-el]}
  \BibitemShut {NoStop}%
\bibitem [{\citenamefont {Kenney}\ \emph {et~al.}(2021)\citenamefont {Kenney},
  \citenamefont {Ortiz}, \citenamefont {Wang}, \citenamefont {Wilson},\ and\
  \citenamefont {Graf}}]{Kenney2021}%
  \BibitemOpen
  \bibfield  {author} {\bibinfo {author} {\bibfnamefont {E.~M.}\ \bibnamefont
  {Kenney}}, \bibinfo {author} {\bibfnamefont {B.~R.}\ \bibnamefont {Ortiz}},
  \bibinfo {author} {\bibfnamefont {C.}~\bibnamefont {Wang}}, \bibinfo {author}
  {\bibfnamefont {S.~D.}\ \bibnamefont {Wilson}},\ and\ \bibinfo {author}
  {\bibfnamefont {M.~J.}\ \bibnamefont {Graf}},\ }\bibfield  {title} {\bibinfo
  {title} {{Absence of local moments in the kagome metal KV$_3$Sb$_5$ as
  determined by muon spin spectroscopy}},\ }\href
  {https://doi.org/10.1088/1361-648x/abe8f9} {\bibfield  {journal} {\bibinfo
  {journal} {Journal of Physics: Condensed Matter}\ }\textbf {\bibinfo {volume}
  {33}},\ \bibinfo {pages} {235801} (\bibinfo {year} {2021})}\BibitemShut
  {NoStop}%
\bibitem [{\citenamefont {Yang}\ \emph {et~al.}(2020)\citenamefont {Yang},
  \citenamefont {Wang}, \citenamefont {Ortiz}, \citenamefont {Liu},
  \citenamefont {Gayles}, \citenamefont {Derunova}, \citenamefont
  {Gonzalez-Hernandez}, \citenamefont {Šmejkal}, \citenamefont {Chen},
  \citenamefont {Parkin}, \citenamefont {Wilson}, \citenamefont {Toberer},
  \citenamefont {McQueen},\ and\ \citenamefont {Ali}}]{Yang2020}%
  \BibitemOpen
  \bibfield  {author} {\bibinfo {author} {\bibfnamefont {S.-Y.}\ \bibnamefont
  {Yang}}, \bibinfo {author} {\bibfnamefont {Y.}~\bibnamefont {Wang}}, \bibinfo
  {author} {\bibfnamefont {B.~R.}\ \bibnamefont {Ortiz}}, \bibinfo {author}
  {\bibfnamefont {D.}~\bibnamefont {Liu}}, \bibinfo {author} {\bibfnamefont
  {J.}~\bibnamefont {Gayles}}, \bibinfo {author} {\bibfnamefont
  {E.}~\bibnamefont {Derunova}}, \bibinfo {author} {\bibfnamefont
  {R.}~\bibnamefont {Gonzalez-Hernandez}}, \bibinfo {author} {\bibfnamefont
  {L.}~\bibnamefont {Šmejkal}}, \bibinfo {author} {\bibfnamefont
  {Y.}~\bibnamefont {Chen}}, \bibinfo {author} {\bibfnamefont {S.~S.~P.}\
  \bibnamefont {Parkin}}, \bibinfo {author} {\bibfnamefont {S.~D.}\
  \bibnamefont {Wilson}}, \bibinfo {author} {\bibfnamefont {E.~S.}\
  \bibnamefont {Toberer}}, \bibinfo {author} {\bibfnamefont {T.}~\bibnamefont
  {McQueen}},\ and\ \bibinfo {author} {\bibfnamefont {M.~N.}\ \bibnamefont
  {Ali}},\ }\bibfield  {title} {\bibinfo {title} {{Giant, unconventional
  anomalous Hall effect in the metallic frustrated magnet candidate,
  KV$_3$Sb$_5$}},\ }\href {https://doi.org/10.1126/sciadv.abb6003} {\bibfield
  {journal} {\bibinfo  {journal} {Science Advances}\ }\textbf {\bibinfo
  {volume} {6}},\ \bibinfo {pages} {eabb6003} (\bibinfo {year}
  {2020})}\BibitemShut {NoStop}%
\bibitem [{\citenamefont {Yu}\ \emph {et~al.}(2021{\natexlab{a}})\citenamefont
  {Yu}, \citenamefont {Wu}, \citenamefont {Wang}, \citenamefont {Lei},
  \citenamefont {Zhuo}, \citenamefont {Ying},\ and\ \citenamefont
  {Chen}}]{Yu2021a}%
  \BibitemOpen
  \bibfield  {author} {\bibinfo {author} {\bibfnamefont {F.~H.}\ \bibnamefont
  {Yu}}, \bibinfo {author} {\bibfnamefont {T.}~\bibnamefont {Wu}}, \bibinfo
  {author} {\bibfnamefont {Z.~Y.}\ \bibnamefont {Wang}}, \bibinfo {author}
  {\bibfnamefont {B.}~\bibnamefont {Lei}}, \bibinfo {author} {\bibfnamefont
  {W.~Z.}\ \bibnamefont {Zhuo}}, \bibinfo {author} {\bibfnamefont {J.~J.}\
  \bibnamefont {Ying}},\ and\ \bibinfo {author} {\bibfnamefont {X.~H.}\
  \bibnamefont {Chen}},\ }\bibfield  {title} {\bibinfo {title} {Concurrence of
  anomalous hall effect and charge density wave in a superconducting
  topological kagome metal},\ }\href
  {https://doi.org/10.1103/PhysRevB.104.L041103} {\bibfield  {journal}
  {\bibinfo  {journal} {Phys. Rev. B}\ }\textbf {\bibinfo {volume} {104}},\
  \bibinfo {pages} {L041103} (\bibinfo {year}
  {2021}{\natexlab{a}})}\BibitemShut {NoStop}%
\bibitem [{\citenamefont {Fradkin}\ \emph {et~al.}(2015)\citenamefont
  {Fradkin}, \citenamefont {Kivelson},\ and\ \citenamefont
  {Tranquada}}]{Fradkin2015}%
  \BibitemOpen
  \bibfield  {author} {\bibinfo {author} {\bibfnamefont {E.}~\bibnamefont
  {Fradkin}}, \bibinfo {author} {\bibfnamefont {S.~A.}\ \bibnamefont
  {Kivelson}},\ and\ \bibinfo {author} {\bibfnamefont {J.~M.}\ \bibnamefont
  {Tranquada}},\ }\bibfield  {title} {\bibinfo {title} {Colloquium: Theory of
  intertwined orders in high temperature superconductors},\ }\href
  {https://doi.org/10.1103/RevModPhys.87.457} {\bibfield  {journal} {\bibinfo
  {journal} {Rev. Mod. Phys.}\ }\textbf {\bibinfo {volume} {87}},\ \bibinfo
  {pages} {457} (\bibinfo {year} {2015})}\BibitemShut {NoStop}%
\bibitem [{\citenamefont {Dai}\ \emph {et~al.}(2018)\citenamefont {Dai},
  \citenamefont {Zhang}, \citenamefont {Senthil},\ and\ \citenamefont
  {Lee}}]{Dai2018}%
  \BibitemOpen
  \bibfield  {author} {\bibinfo {author} {\bibfnamefont {Z.}~\bibnamefont
  {Dai}}, \bibinfo {author} {\bibfnamefont {Y.-H.}\ \bibnamefont {Zhang}},
  \bibinfo {author} {\bibfnamefont {T.}~\bibnamefont {Senthil}},\ and\ \bibinfo
  {author} {\bibfnamefont {P.~A.}\ \bibnamefont {Lee}},\ }\bibfield  {title}
  {\bibinfo {title} {Pair-density waves, charge-density waves, and vortices in
  high-${T}_{c}$ cuprates},\ }\href
  {https://doi.org/10.1103/PhysRevB.97.174511} {\bibfield  {journal} {\bibinfo
  {journal} {Phys. Rev. B}\ }\textbf {\bibinfo {volume} {97}},\ \bibinfo
  {pages} {174511} (\bibinfo {year} {2018})}\BibitemShut {NoStop}%
\bibitem [{\citenamefont {Chen}\ \emph
  {et~al.}(2021{\natexlab{b}})\citenamefont {Chen}, \citenamefont {Wang},
  \citenamefont {Yin}, \citenamefont {Gu}, \citenamefont {Jiang}, \citenamefont
  {Tu}, \citenamefont {Gong}, \citenamefont {Uwatoko}, \citenamefont {Sun},
  \citenamefont {Lei}, \citenamefont {Hu},\ and\ \citenamefont
  {Cheng}}]{Chen2021}%
  \BibitemOpen
  \bibfield  {author} {\bibinfo {author} {\bibfnamefont {K.~Y.}\ \bibnamefont
  {Chen}}, \bibinfo {author} {\bibfnamefont {N.~N.}\ \bibnamefont {Wang}},
  \bibinfo {author} {\bibfnamefont {Q.~W.}\ \bibnamefont {Yin}}, \bibinfo
  {author} {\bibfnamefont {Y.~H.}\ \bibnamefont {Gu}}, \bibinfo {author}
  {\bibfnamefont {K.}~\bibnamefont {Jiang}}, \bibinfo {author} {\bibfnamefont
  {Z.~J.}\ \bibnamefont {Tu}}, \bibinfo {author} {\bibfnamefont {C.~S.}\
  \bibnamefont {Gong}}, \bibinfo {author} {\bibfnamefont {Y.}~\bibnamefont
  {Uwatoko}}, \bibinfo {author} {\bibfnamefont {J.~P.}\ \bibnamefont {Sun}},
  \bibinfo {author} {\bibfnamefont {H.~C.}\ \bibnamefont {Lei}}, \bibinfo
  {author} {\bibfnamefont {J.~P.}\ \bibnamefont {Hu}},\ and\ \bibinfo {author}
  {\bibfnamefont {J.~G.}\ \bibnamefont {Cheng}},\ }\bibfield  {title} {\bibinfo
  {title} {{Double superconducting dome and triple enhancement of ${T}_{c}$ in
  the kagome superconductor CsV$_{3}$Sb$_{5}$ under high pressure}},\ }\href
  {https://doi.org/10.1103/PhysRevLett.126.247001} {\bibfield  {journal}
  {\bibinfo  {journal} {Phys. Rev. Lett.}\ }\textbf {\bibinfo {volume} {126}},\
  \bibinfo {pages} {247001} (\bibinfo {year} {2021}{\natexlab{b}})}\BibitemShut
  {NoStop}%
\bibitem [{\citenamefont {Du}\ \emph {et~al.}(2021)\citenamefont {Du},
  \citenamefont {Luo}, \citenamefont {Ortiz}, \citenamefont {Chen},
  \citenamefont {Duan}, \citenamefont {Zhang}, \citenamefont {Lu},
  \citenamefont {Wilson}, \citenamefont {Song},\ and\ \citenamefont
  {Yuan}}]{Du2021}%
  \BibitemOpen
  \bibfield  {author} {\bibinfo {author} {\bibfnamefont {F.}~\bibnamefont
  {Du}}, \bibinfo {author} {\bibfnamefont {S.}~\bibnamefont {Luo}}, \bibinfo
  {author} {\bibfnamefont {B.~R.}\ \bibnamefont {Ortiz}}, \bibinfo {author}
  {\bibfnamefont {Y.}~\bibnamefont {Chen}}, \bibinfo {author} {\bibfnamefont
  {W.}~\bibnamefont {Duan}}, \bibinfo {author} {\bibfnamefont {D.}~\bibnamefont
  {Zhang}}, \bibinfo {author} {\bibfnamefont {X.}~\bibnamefont {Lu}}, \bibinfo
  {author} {\bibfnamefont {S.~D.}\ \bibnamefont {Wilson}}, \bibinfo {author}
  {\bibfnamefont {Y.}~\bibnamefont {Song}},\ and\ \bibinfo {author}
  {\bibfnamefont {H.}~\bibnamefont {Yuan}},\ }\bibfield  {title} {\bibinfo
  {title} {{Pressure-induced double superconducting domes and charge
  instability in the kagome metal KV$_3$Sb$_5$}},\ }\href
  {https://doi.org/10.1103/PhysRevB.103.L220504} {\bibfield  {journal}
  {\bibinfo  {journal} {Phys. Rev. B}\ }\textbf {\bibinfo {volume} {103}},\
  \bibinfo {pages} {L220504} (\bibinfo {year} {2021})}\BibitemShut {NoStop}%
\bibitem [{\citenamefont {Yu}\ \emph {et~al.}(2021{\natexlab{b}})\citenamefont
  {Yu}, \citenamefont {Ma}, \citenamefont {Zhuo}, \citenamefont {Liu},
  \citenamefont {Wen}, \citenamefont {Lei}, \citenamefont {Ying},\ and\
  \citenamefont {Chen}}]{Yu2021}%
  \BibitemOpen
  \bibfield  {author} {\bibinfo {author} {\bibfnamefont {F.~H.}\ \bibnamefont
  {Yu}}, \bibinfo {author} {\bibfnamefont {D.~H.}\ \bibnamefont {Ma}}, \bibinfo
  {author} {\bibfnamefont {W.~Z.}\ \bibnamefont {Zhuo}}, \bibinfo {author}
  {\bibfnamefont {S.~Q.}\ \bibnamefont {Liu}}, \bibinfo {author} {\bibfnamefont
  {X.~K.}\ \bibnamefont {Wen}}, \bibinfo {author} {\bibfnamefont
  {B.}~\bibnamefont {Lei}}, \bibinfo {author} {\bibfnamefont {J.~J.}\
  \bibnamefont {Ying}},\ and\ \bibinfo {author} {\bibfnamefont {X.~H.}\
  \bibnamefont {Chen}},\ }\bibfield  {title} {\bibinfo {title} {{Unusual
  competition of superconductivity and charge-density-wave state in a
  compressed topological kagome metal}},\ }\href
  {https://doi.org/10.1038/s41467-021-23928-w} {\bibfield  {journal} {\bibinfo
  {journal} {Nature Communications}\ }\textbf {\bibinfo {volume} {12}},\
  \bibinfo {pages} {3645} (\bibinfo {year} {2021}{\natexlab{b}})}\BibitemShut
  {NoStop}%
\bibitem [{\citenamefont {Wilson}(2021)}]{Wilson2021}%
  \BibitemOpen
  \bibfield  {author} {\bibinfo {author} {\bibfnamefont {S.~D.}\ \bibnamefont
  {Wilson}},\ }\bibfield  {title} {\bibinfo {title} {Competition under
  pressure},\ }\href {https://doi.org/10.1038/s41567-021-01319-8} {\bibfield
  {journal} {\bibinfo  {journal} {Nature Physics}\ }\textbf {\bibinfo {volume}
  {17}},\ \bibinfo {pages} {885} (\bibinfo {year} {2021})}\BibitemShut
  {NoStop}%
\bibitem [{\citenamefont {Tan}\ \emph {et~al.}(2021)\citenamefont {Tan},
  \citenamefont {Liu}, \citenamefont {Wang},\ and\ \citenamefont
  {Yan}}]{Tan2021}%
  \BibitemOpen
  \bibfield  {author} {\bibinfo {author} {\bibfnamefont {H.}~\bibnamefont
  {Tan}}, \bibinfo {author} {\bibfnamefont {Y.}~\bibnamefont {Liu}}, \bibinfo
  {author} {\bibfnamefont {Z.}~\bibnamefont {Wang}},\ and\ \bibinfo {author}
  {\bibfnamefont {B.}~\bibnamefont {Yan}},\ }\bibfield  {title} {\bibinfo
  {title} {Charge density waves and electronic properties of superconducting
  kagome metals},\ }\href {https://doi.org/10.1103/PhysRevLett.127.046401}
  {\bibfield  {journal} {\bibinfo  {journal} {Phys. Rev. Lett.}\ }\textbf
  {\bibinfo {volume} {127}},\ \bibinfo {pages} {046401} (\bibinfo {year}
  {2021})}\BibitemShut {NoStop}%
\bibitem [{SM()}]{SM}%
  \BibitemOpen
  \href@noop {} {}\bibinfo {note} {See Supplementary Materials for
  thermodynamic, transport and ARPES measurement methods, the calculation
  methods and the supercell calculations of Ti-doped
  CsV$_3$Sb$_5$.}\BibitemShut {Stop}%
\bibitem [{\citenamefont {Wu}\ \emph {et~al.}(2021)\citenamefont {Wu},
  \citenamefont {Schwemmer}, \citenamefont {M\"uller}, \citenamefont
  {Consiglio}, \citenamefont {Sangiovanni}, \citenamefont {Di~Sante},
  \citenamefont {Iqbal}, \citenamefont {Hanke}, \citenamefont {Schnyder},
  \citenamefont {Denner}, \citenamefont {Fischer}, \citenamefont {Neupert},\
  and\ \citenamefont {Thomale}}]{Wu2021}%
  \BibitemOpen
  \bibfield  {author} {\bibinfo {author} {\bibfnamefont {X.}~\bibnamefont
  {Wu}}, \bibinfo {author} {\bibfnamefont {T.}~\bibnamefont {Schwemmer}},
  \bibinfo {author} {\bibfnamefont {T.}~\bibnamefont {M\"uller}}, \bibinfo
  {author} {\bibfnamefont {A.}~\bibnamefont {Consiglio}}, \bibinfo {author}
  {\bibfnamefont {G.}~\bibnamefont {Sangiovanni}}, \bibinfo {author}
  {\bibfnamefont {D.}~\bibnamefont {Di~Sante}}, \bibinfo {author}
  {\bibfnamefont {Y.}~\bibnamefont {Iqbal}}, \bibinfo {author} {\bibfnamefont
  {W.}~\bibnamefont {Hanke}}, \bibinfo {author} {\bibfnamefont {A.~P.}\
  \bibnamefont {Schnyder}}, \bibinfo {author} {\bibfnamefont {M.~M.}\
  \bibnamefont {Denner}}, \bibinfo {author} {\bibfnamefont {M.~H.}\
  \bibnamefont {Fischer}}, \bibinfo {author} {\bibfnamefont {T.}~\bibnamefont
  {Neupert}},\ and\ \bibinfo {author} {\bibfnamefont {R.}~\bibnamefont
  {Thomale}},\ }\bibfield  {title} {\bibinfo {title} {Nature of unconventional
  pairing in the kagome superconductors {$A$Cs$_3$Sb$_5$}
  ({$A=\mathrm{K},\mathrm{Rb},\mathrm{Cs}$})},\ }\href
  {https://doi.org/10.1103/PhysRevLett.127.177001} {\bibfield  {journal}
  {\bibinfo  {journal} {Phys. Rev. Lett.}\ }\textbf {\bibinfo {volume} {127}},\
  \bibinfo {pages} {177001} (\bibinfo {year} {2021})}\BibitemShut {NoStop}%
\bibitem [{\citenamefont {Mu}\ \emph {et~al.}(2021)\citenamefont {Mu},
  \citenamefont {Yin}, \citenamefont {Tu}, \citenamefont {Gong}, \citenamefont
  {Lei}, \citenamefont {Li},\ and\ \citenamefont {Luo}}]{Mu2021}%
  \BibitemOpen
  \bibfield  {author} {\bibinfo {author} {\bibfnamefont {C.}~\bibnamefont
  {Mu}}, \bibinfo {author} {\bibfnamefont {Q.}~\bibnamefont {Yin}}, \bibinfo
  {author} {\bibfnamefont {Z.}~\bibnamefont {Tu}}, \bibinfo {author}
  {\bibfnamefont {C.}~\bibnamefont {Gong}}, \bibinfo {author} {\bibfnamefont
  {H.}~\bibnamefont {Lei}}, \bibinfo {author} {\bibfnamefont {Z.}~\bibnamefont
  {Li}},\ and\ \bibinfo {author} {\bibfnamefont {J.}~\bibnamefont {Luo}},\
  }\bibfield  {title} {\bibinfo {title} {S-wave superconductivity in kagome
  metal {CsV}$_3${Sb}$_5$ revealed by $^{121/123}${Sb} {NQR} and $^{51}${V}
  {NMR} measurements},\ }\href {https://doi.org/10.1088/0256-307x/38/7/077402}
  {\bibfield  {journal} {\bibinfo  {journal} {Chinese Physics Letters}\
  }\textbf {\bibinfo {volume} {38}},\ \bibinfo {pages} {077402} (\bibinfo
  {year} {2021})}\BibitemShut {NoStop}%
\bibitem [{\citenamefont {Duan}\ \emph {et~al.}(2021)\citenamefont {Duan},
  \citenamefont {Nie}, \citenamefont {Luo}, \citenamefont {Yu}, \citenamefont
  {Ortiz}, \citenamefont {Yin}, \citenamefont {Su}, \citenamefont {Du},
  \citenamefont {Wang}, \citenamefont {Chen}, \citenamefont {Lu}, \citenamefont
  {Ying}, \citenamefont {Wilson}, \citenamefont {Chen}, \citenamefont {Song},\
  and\ \citenamefont {Yuan}}]{Duan2021}%
  \BibitemOpen
  \bibfield  {author} {\bibinfo {author} {\bibfnamefont {W.}~\bibnamefont
  {Duan}}, \bibinfo {author} {\bibfnamefont {Z.}~\bibnamefont {Nie}}, \bibinfo
  {author} {\bibfnamefont {S.}~\bibnamefont {Luo}}, \bibinfo {author}
  {\bibfnamefont {F.}~\bibnamefont {Yu}}, \bibinfo {author} {\bibfnamefont
  {B.~R.}\ \bibnamefont {Ortiz}}, \bibinfo {author} {\bibfnamefont
  {L.}~\bibnamefont {Yin}}, \bibinfo {author} {\bibfnamefont {H.}~\bibnamefont
  {Su}}, \bibinfo {author} {\bibfnamefont {F.}~\bibnamefont {Du}}, \bibinfo
  {author} {\bibfnamefont {A.}~\bibnamefont {Wang}}, \bibinfo {author}
  {\bibfnamefont {Y.}~\bibnamefont {Chen}}, \bibinfo {author} {\bibfnamefont
  {X.}~\bibnamefont {Lu}}, \bibinfo {author} {\bibfnamefont {J.}~\bibnamefont
  {Ying}}, \bibinfo {author} {\bibfnamefont {S.~D.}\ \bibnamefont {Wilson}},
  \bibinfo {author} {\bibfnamefont {X.}~\bibnamefont {Chen}}, \bibinfo {author}
  {\bibfnamefont {Y.}~\bibnamefont {Song}},\ and\ \bibinfo {author}
  {\bibfnamefont {H.}~\bibnamefont {Yuan}},\ }\bibfield  {title} {\bibinfo
  {title} {Nodeless superconductivity in the kagome metal {CsV}$_3${Sb}$_5$},\
  }\href {https://doi.org/10.1007/s11433-021-1747-7} {\bibfield  {journal}
  {\bibinfo  {journal} {Science China Physics, Mechanics \& Astronomy}\
  }\textbf {\bibinfo {volume} {64}},\ \bibinfo {pages} {107462} (\bibinfo
  {year} {2021})}\BibitemShut {NoStop}%
\end{thebibliography}%
\end{document}